\documentclass[10pt,aps,prd,twocolumn,showpacs,amsmath,nofootinbib,floatfix]{revtex4-2}

\usepackage[utf8]{inputenc}
\usepackage[american]{babel}
\usepackage[a4paper,margin=1.6cm]{geometry}

\usepackage{calrsfs}
\DeclareMathAlphabet{\pazocal}{OMS}{zplm}{m}{n}

\usepackage{amsmath,amssymb,mathtools}
\usepackage{booktabs}
\usepackage{multirow}
\usepackage{graphicx}
\usepackage[labelfont=bf,font+=smaller,justification=centerlast]{caption}
\usepackage[labelfont=normal,font=normal,font=small]{subcaption}
\usepackage{verbatim}

\usepackage{hyperref}
\usepackage{cleveref}
\renewcommand{\cref}[1]{\Cref{#1}}

\Crefname{equation}{Eq.}{Eqs.}
\Crefname{figure}{Fig.}{Figs.}
\Crefname{tabular}{Tab.}{Tabs.}
\Crefname{section}{Sec.}{Secs.}

\hypersetup{%
colorlinks,%
citecolor=blue,urlcolor=blue,linkcolor=blue,%
hypertexnames=false%
}
\makeatletter
\renewcommand\HyPsd@CatcodeWarning[1]{}     
\makeatother

\allowdisplaybreaks

\hypersetup{%
pdftitle={K -> pi g* g* transitions at leading order and beyond},%
pdfauthor={T. Husek}%
}

\def\KL{K_\text{L}}
\def\KS{K_\text{S}}
\def\ChPT{\Lambda_\chi}

\def\La{\mathcal{L}}
\def\Lb{\pazocal{L}}
\def\cA{c^{(6)}}
\def\cB{d^{(6)}}
\def\cBr{d^{(6)\text{r}}}
\def\cc{c}
\def\GF{G_\text{F}}
\def\go{\widetilde g_8}

\def\w{w}
\def\d{\mathrm{d}}
\def\Jbar{\bar J}
\def\Jtilde{\widetilde J}

\def\qEP{q^{(\text{E},+)}}
\def\qEM{q^{(\text{E},-)}}
\def\qIP{q^{(\text{I},+)}}
\def\qIM{q^{(\text{I},-)}}
\def\qEPM{q^{(\text{E},\pm)}}
\def\qIPM{q^{(\text{I},\pm)}}

\begin{document}

\onecolumngrid
\thispagestyle{empty}

\ \vspace{2cm}

\begin{center}
{\LARGE\bf $K\to\pi\gamma^*\gamma^*$ transitions at leading order and beyond}\\[15mm]
{\large\bf Tom\'{a}\v{s} Husek}\\[3mm]

{Institute of Particle and Nuclear Physics, Charles University, V Hole\v sovi\v ck\'ach 2, 180 00 Prague, Czech Republic}\\
{School of Physics and Astronomy, University of Birmingham, Edgbaston, Birmingham, B15 2TT, United Kingdom}\\[15mm]

{\large\bf Abstract}
\end{center}

The transition amplitude of a kaon to a pion and two off-shell photons is studied. First, it is computed at leading order (one-loop level) of the Chiral Perturbation Theory expansion. Explicit analytical results for the leading-order amplitude are presented, constituting the first complete calculation for the doubly off-shell case. Subsequently, it is reevaluated by employing a refined diagrammatic notation and a generic ansatz incorporating effects beyond leading order. The dependence on the underlying $K\pi PP$ vertex parameters is analyzed. This offers valuable insights into amplitude properties and allows inclusion of unitarity corrections from $K\to3\pi$, yielding the complete $K\to\pi\gamma^*\gamma^*$ amplitude structure. Both the charged and neutral channels are treated in parallel. The presented results provide crucial input for phenomenological studies of related rare decays like $K\to\pi\ell^+\ell^-[\gamma]$ or $K\to\pi\ell_1^+\ell_1^-\ell_2^+\ell_2^-$ and support ongoing precision measurements at experiments like NA62 at CERN. These results may also find application in other related processes, including $\eta^{(\prime)}$ decays.

\vfill

\hspace{2.7cm}{\large\bf CONTENTS}\\[2mm]

\twocolumngrid
{
\makeatletter
\def\tocname{\vspace{-9.15mm}}
\makeatother
\tableofcontents
}
\onecolumngrid

\vfill

\setcounter{page}{0}
\newpage

\title{$K\to\pi\gamma^*\gamma^*$ transitions at leading order and beyond}
\author{Tom\'{a}\v{s} Husek}
\email{tomas.husek@matfyz.cuni.cz; t.husek@bham.ac.uk}
\affiliation{Institute of Particle and Nuclear Physics, Charles University, V Hole\v sovi\v ck\'ach 2, 180 00 Prague, Czech Republic}
\affiliation{\vspace{1mm}School of Physics and Astronomy, University of Birmingham, Edgbaston, Birmingham, B15 2TT, United Kingdom}
\date{October 23, 2025}

\begin{abstract}
The transition amplitude of a kaon to a pion and two off-shell photons is studied. First, it is computed at leading order (one-loop level) of the Chiral Perturbation Theory expansion. Explicit analytical results for the leading-order amplitude are presented, constituting the first complete calculation for the doubly off-shell case. Subsequently, it is reevaluated by employing a refined diagrammatic notation and a generic ansatz incorporating effects beyond leading order. The dependence on the underlying $K\pi PP$ vertex parameters is analyzed. This offers valuable insights into amplitude properties and allows inclusion of unitarity corrections from $K\to3\pi$, yielding the complete $K\to\pi\gamma^*\gamma^*$ amplitude structure. Both the charged and neutral channels are treated in parallel. The presented results provide crucial input for phenomenological studies of related rare decays like $K\to\pi\ell^+\ell^-[\gamma]$ or $K\to\pi\ell_1^+\ell_1^-\ell_2^+\ell_2^-$ and support ongoing precision measurements at experiments like NA62 at CERN. These results may also find application in other related processes, including $\eta^{(\prime)}$ decays.
\end{abstract}


\maketitle

\section{Introduction}

Despite CERN's decision to discontinue its prominent kaon program, there are still years of data taking and analysis by the NA62 Collaboration ahead:
The NA62 experiment located at CERN's North Area has collected samples of over $10^{13}$ $K^+$ decays to both the dielectron and dimuon final states, and further data are to be collected in 2025 and 2026.
The ongoing NA62 physics program, based on these datasets and motivating this work, includes
\begin{enumerate}
    \item[1)] precision lepton-flavor universality tests via measurements of the form factors and branching ratios of $K^+\to\pi^+\ell^+\ell^-$ decays~\cite{NA62:2022qes},
    \item[2)] studies of the $K^+\to\pi^+\gamma\ell^+\ell^-$ decays, and
    \item[3)] searches for the $K^+\to\pi^+\ell_1^+\ell_1^-\ell_2^+\ell_2^-$ decays~\cite{NA62:2023rvm}.
\end{enumerate}
Expected single-event sensitivities for branching ratios of certain rare decays are as low as $10^{-12}$.

The mentioned processes share an underlying physics case: nonleptonic strangeness-changing neutral-current weak transitions $K\to\pi\gamma^{(*)}[\gamma^{(*)}]$.
A suitable framework for describing these long-distance-dominated conversions is the Chiral Perturbation Theory (ChPT), extended by both weak and electromagnetic (EM) perturbations.
Such a framework was worked out in Refs.~\cite{Ecker:1987qi,Ecker:1987hd}.
In these and a series of works that followed, the \mbox{$K\to\pi\gamma^*$} and \mbox{$K\to\pi\gamma^{(*)}\gamma$} transitions were studied.
In particular, this was in the context of the decays 
\begin{center}
\setlength{\tabcolsep}{10pt}
\renewcommand{\arraystretch}{1.2}
\begin{tabular}{l l}
    $\KS\to\pi^0\ell^+\ell^-$~\cite{Donoghue:1994yt,DAmbrosio:1998gur}\,,
  & $K^+\to\pi^+\ell^+\ell^-$~\cite{DAmbrosio:1998gur}\,,\\
    $\KL\to\pi^0\gamma\gamma$~\cite{Ecker:1987fm,Cappiello:1992kk,Cohen:1993ta}\,,
  & $K^+\to\pi^+\gamma\gamma$~\cite{DAmbrosio:1996cak}\,,\\
    $\KL\to\pi^0\gamma\ell^+\ell^-$~\cite{Donoghue:1997rr,Donoghue:1998ur}\,,
  & $K^+\to\pi^+\gamma\ell^+\ell^-$~\cite{Gabbiani:1998tj}\,.
\end{tabular}
\end{center}

In this paper, I proceed further along the same path by presenting the results for the \mbox{$\KL\to\pi^0\gamma^{*}\gamma^{*}$} and \mbox{$K^+\to\pi^+\gamma^{*}\gamma^{*}$} transitions; i.e., I add the case with two off-shell photons:
The results found in the related literature (like the works listed above) are special cases with one or two vanishing photon virtualities.
I am unaware of a systematic computation of the doubly off-shell case in the literature, even at leading order (LO) in ChPT.
However, restricting the analysis solely to this case would be insufficient.
It was demonstrated in several cases that the LO ChPT calculation is not sufficient when it comes to these weak radiative transitions.
Either the measured branching ratios tend to be significantly larger than the theoretical predictions at LO~\cite{Cappiello:1992kk,Cohen:1993ta}, or the predicted spectral shapes do not describe the data sufficiently well~\cite{NA62:2023olg}.
A leap in the right direction was to include unitarity corrections from $K\to3\pi$ data, which effectively accounts for $\pi\pi$ rescattering before their internal conversion into photon(s)~\cite{Cappiello:1992kk,Cohen:1993ta,DAmbrosio:1996cak,DAmbrosio:1998gur,Donoghue:1997rr,Donoghue:1998ur,Gabbiani:1998tj}.
I have therefore decided to include this correction from the outset.
Unfortunately, as a trade-off, this leads to some obvious complications and makes the expressions for these corrections significantly less elegant and compact than the corresponding LO expression.
Nevertheless, it is still worthwhile to set them out in full here, and the additional subamplitudes due to the unitarity corrections for the doubly off-shell case are attached as \cref{sec:Ai}.

After the main LO ChPT result is presented in \cref{sec:2g}, a rather general formalism is introduced, which offers valuable insights and potentially allows for the direct application of the results in other channels, e.g., decays of $\eta^{(\prime)}$ mesons.
Although I use standard Feynman diagrams first, starting from \cref{sec:general}, I introduce a more suitable diagram notation for radiative vertices, which allows for a direct separation of contributions into gauge-invariant subsets.
The advantages are manifest, and a technical explanation is provided in \cref{sec:notation}.

I first present the effective theoretical (ChPT) framework in \cref{sec:intro} and set the stage by recalculating some results for the one-photon transition form factors at LO in ChPT in \cref{sec:1g}.
This serves both as a cross-check and preparation of a building block used further in two-photon transitions at LO, the calculation of which I describe in \cref{sec:2g}.
The results are presented in a compact form, and expressions for special cases in which one or both photons become on-shell are shown.
In \cref{sec:general}, the LO amplitude is generalized, which allows for understanding some properties of the amplitudes in question, and the ingredients for the NLO unitarity corrections are prepared.
I then repeat the calculation using this generic ansatz and present results beyond LO in \cref{sec:gen_res_NLO}.
This is followed by a discussion in \cref{summary}.
There are several appendices to show further explicit results, support statements, and elaborate on some concepts.
I describe my diagrammatic notation that is used beyond \cref{sec:2g} in \cref{sec:notation}, list the utilized loop functions in \cref{sec:functions}, present the expanded ChPT Lagrangian to LO in \cref{sec:Lagrangians} and the derived amplitudes in \cref{sec:amplitudes}, discuss the Lorentz structure of the $K\to\pi\gamma^*\gamma^*$ amplitude in \cref{sec:structures} and the counterterms (CTs) in \cref{sec:CTs}, show an instructive effective-field-theory (EFT) Lagrangian in \cref{sec:EFT}, and finally, list the cumbersome subamplitudes beyond LO in \cref{sec:Ai}.

\section{Theoretical framework}
\label{sec:intro}

Following Ref.~\cite{Ecker:1987hd}, the (effective) Lagrangian leading to the necessary ingredients for the LO calculation of the electromagnetically induced (weak) kaon decays in question can be expressed in the following way:
\begin{equation}
\begin{aligned}
\Lb
=\Lb_\text{strong}^\text{(+EM)}+\Lb_{\Delta S=1}^\text{(+EM)}+\Lb_\text{WZW}\,.
\end{aligned}
\label{eq:L_ChPT}
\end{equation}
The $\mathcal{O}(p^2)$ strong interactions among the octet of pseudoscalar mesons, enriched with (minimal) EM interactions and including necessary CTs, are contained in $\Lb_\text{strong}^\text{(+EM)}$.
The relevant terms can be written as
\begin{equation}
\begin{aligned}
&\Lb_\text{strong}^\text{(+EM)}
\supset-\frac14F_{\mu\nu}F^{\mu\nu}\\
&+\frac{F_0^2}{4}\Bigl[
\operatorname{Tr}\bigl(D_\mu U(D^\mu U)^\dag\bigr)
+2B_0\!\operatorname{Tr}\bigl(\mathcal{M}U+U^\dag \mathcal{M}\bigr)\Big]\\
&-ieL_9F^{\mu\nu}\operatorname{Tr}\bigl(QD_\mu U(D_\nu U)^\dag+Q(D_\mu U)^\dag D_\nu U\bigr)\\
&+e^2L_{10}F_{\mu\nu}F^{\mu\nu}\operatorname{Tr}\bigl(QUQU^\dag\bigr)\,.
\end{aligned}
\label{eq:L_strong}
\end{equation}
At the same order, the nonleptonic strangeness-changing ($\Delta S=1$) weak interactions in the presence of EM interactions are represented by
\begin{equation}
\begin{aligned}
&\Lb_{\Delta S=1}^\text{(+EM)}
\supset G_8(\La_\mu \La^\mu)_{23}+\text{h.c.}+\text{nonoctet terms}\\
&-\frac{ieG_8}{2F_0^2}F^{\mu\nu}\Big[
w_1\operatorname{Tr}\bigl(Q\lambda_{6-i7}\La_\mu \La_\nu\big)\\
&\qquad\qquad\qquad+w_2\operatorname{Tr}\bigl(Q\La_\mu\lambda_{6-i7}\La_\nu\big)\\
&\qquad\qquad\qquad+w_3\operatorname{Tr}\bigl(Q\La_\mu\big)\operatorname{Tr}\bigl(\lambda_{6-i7}\La_\nu\big)
\Big]+\text{h.c.}\\
&+\frac12e^2G_8F_0^2w_4F_{\mu\nu}F^{\mu\nu}\operatorname{Tr}\bigl(\lambda_{6-i7}QUQU^\dag \bigr)+\text{h.c.},
\end{aligned}
\label{eq:L_DS1}
\end{equation}
with the covariant left-handed current \mbox{$\La_\mu=iF_0^2U(D_\mu U)^\dag$} and with $(\cdots)_{ij}$ selecting a particular matrix entry.
Finally, the relevant contributions of the QCD anomaly are governed by
\begin{equation}
\begin{aligned}
\Lb_\text{WZW}
&=\frac{e^2}{32\pi^2F_0}\,\varepsilon_{\mu\nu\rho\sigma}F^{\mu\nu}F^{\rho\sigma}\Big(\pi^0+\tfrac1{\sqrt{3}}\eta\Big)\,.
\end{aligned}
\end{equation}
Above, $U=U\bigl(\Phi(x)\big)$, $\Phi(x)=\sum_{i=1}^8\lambda_i\phi_i(x)$ is the pseudoscalar matrix, with $\lambda_i$ Gell-Mann matrices and $\phi_i(x)$ eight Goldstone-boson fields, $\mathcal{M}=\operatorname{diag}(m_u,m_d,m_s)$, \mbox{$Q=\frac13\operatorname{diag}(2,-1,-1)$}, and $\lambda_{6-i7}=\lambda_6-i\lambda_7$.
The covariant derivative is defined as $D_\mu P(x)=[\partial_\mu-i\eta_PeA_\mu]P(x)$, with $\eta_P$ being the charge of the pseudoscalar $P$ in the units of $e$, and the EM field-strength tensor is $F_{\mu\nu}=\partial_\mu A_\nu-\partial_\nu A_\mu$.
Regarding the quark masses, they are translated into pseudoscalar masses using
\begin{equation}
B_0
=\frac{M_{\pi^+}^2}{m_u+m_d}
=\frac{M_{K^+}^2}{m_u+m_s}
=\frac{M_{K^0}^2}{m_d+m_s}\,.
\end{equation}
The rest are couplings and low-energy constants: $e=\sqrt{4\pi\alpha}$ and $G_8=\go\GF$, with $\go\equiv g_8\frac1{\sqrt{2}}V_{ud}V_{us}^*$ and $\GF=\frac1{\sqrt{2}v^2}$ the Fermi constant; $v\approx246$\,GeV.
In this work, I do not explicitly treat the terms proportional to $G_{27}$ [nonoctet terms in \cref{eq:L_DS1}].
They can, however, be included by employing the general approach introduced after \cref{sec:2g}.

The covariant kinetic and mass terms quadratic in pseudoscalar fields can be diagonalized, leading to the absence of (off-diagonal) flavor-mixing propagators and $\mathcal{O}(p^2)$ $K\pi\gamma(\gamma)$ vertices.
To the first order in $\GF$, this is achieved by field transformations shown explicitly in Eq.~(2.15) of Ref.~\cite{Ecker:1987hd}:
For instance, in the case of the charged mesons, these read
\begin{alignat}{3}
    \pi^+&\to\pi^+&&-2G_8F_0^2\frac{M_K^2}{M_K^2-M_\pi^2}K^+\,,\notag\\
    K^+&\to K^+&&+2G_8^*F_0^2\frac{M_\pi^2}{M_K^2-M_\pi^2}\,\pi^+\,;
\end{alignat}
transformations for $\pi^-$ and $K^-$ are implied by hermitian conjugation.
Consequently, all weak vertices involve at least three pseudoscalar fields~\cite{Ecker:1987hd}.

To obtain the LO ChPT results shown in \cref{sec:1g,sec:2g}, I used two parametrizations of the Nambu--Goldstone-boson manifold,
\begin{subequations}
\label{eq:params}
\begin{align}
U_1(x)&=\exp\bigl(A(x)\big)\,,\label{eq:param_exp}\\
U_2(x)&=\exp\bigl(\tfrac1{18}\!\operatorname{Tr}\bigl[A^3(x)\big]\big)\Big[A(x)+\sqrt{1+A^2(x)}\Big]\,,\label{eq:param_sqrt}
\end{align}
\end{subequations}
with $A(x)=i\Phi(x)/F_0$:
These definitions guarantee that $\det U(x)=1$.
This serves as a simple cross-check of the computation, as the final results should be, in addition to being finite and gauge invariant, also parametrization independent.

The notation and convention for the UV-divergent and finite parts of the low-energy constants (LECs) appearing (not only) in the CTs of \cref{eq:L_strong,eq:L_DS1} is
\begin{equation}
L_i
=-\frac{\Gamma(L_i)}2\,\kappa\,\frac1{\tilde\epsilon}+L_i^{\text{r}}\,,
\end{equation}
with $\kappa\equiv1/{(4\pi)^2}$, $L_i^{\text{r}}\equiv L_i^{\text{r}}(\mu)$, $\tilde\epsilon\equiv\tilde\epsilon(\mu)$, and
\begin{equation}
\frac1{\tilde\epsilon}
\equiv\frac{1}{\epsilon}-\gamma_\text{E}+\log4\pi-\log\mu^2+1\,,\qquad
\epsilon=2-\frac d2\,.
\label{eq:epstilde}
\end{equation}
(Notice the additional `+1' with respect to the $\overline{\text{MS}}$ scheme.)
I also employ
\begin{equation}
    L_P\equiv L_P(\mu)\equiv\kappa\log\frac{M_P^2}{\mu^2}\,.
\end{equation}

\section{One-photon transition}
\label{sec:1g}

The computation of the one-photon transition \mbox{$K\to\pi\gamma^*$} is a natural first step, as it will later serve as a building block for a part of the charged-channel two-photon-transition amplitude, specifically for the one-particle-reducible (1PR) diagrams in the sense of \cref{fig:K->pigg}, discussed at the beginning of \cref{sec:2g}.
Having this in mind and for consistency checks, the momenta of the mesons are kept off-shell during the calculation.
I will treat the charged ($K^+$) and neutral ($K^0$) channels in parallel.

\begin{figure}[t]
\centering
\includegraphics[width=0.98\columnwidth]{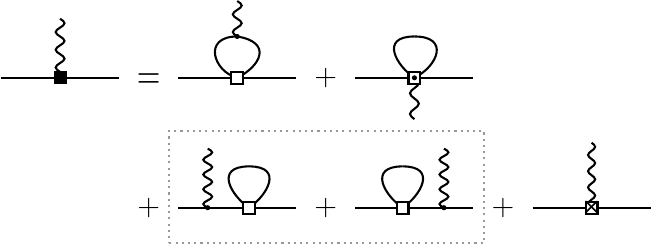}
\caption{
    The representation of the $K\to\pi\gamma^*$ transition amplitude in terms of Feynman diagrams emerging from the ChPT Lagrangian~\eqref{eq:L_ChPT}.
    The minimal electromagnetic coupling is denoted by a small dot and the weak coupling \big($K\pi\pi\pi$ or $K\pi KK$\big) by an empty square; the combination of both is naturally represented by a square with a dot.
    Consequently, there are both (charged) pions and kaons in the loops; the contributions with neutral-meson loops cancel out at the considered orders.
    The counterterm contributions are represented by a square with a cross.
    The two one-particle-reducible (1PR) diagrams (those with photons attached to external legs) are absent in the neutral channel and placed within the gray dotted box.
}
\label{fig:K->pig_LO_ChPT_a}
\end{figure}

Employing the diagonal basis~\cite{Ecker:1987hd}, the diagrams associated with all the contributions appearing at $\mathcal{O}(p^4)$ (LO) in ChPT are shown in \cref{fig:K->pig_LO_ChPT_a}.
%
%
The intermediate result with off-shell meson legs can be written as
\begin{multline}
\mathcal{M}_\rho^{(4)}\big(K(P)\to\pi(r)\gamma_\rho^*(k)\big)\\
=eF(k^2)
\big[
k^2r_\rho
-(k\cdot r)k_\rho
\big]\\
+e\widetilde F(k^2)(M_K^2+M_\pi^2)(P+r)_\rho\,.
\end{multline}
The form factors $F(q^2)$ and $\widetilde F(q^2)$ are expressed below.

The off-shell-emerging part of the charged-channel amplitude depends on the Goldstone-boson-manifold parametrization, and it can be written, employing the two considered parametrizations~\eqref{eq:params}, as
\begin{multline}
\widetilde F^{(+)}(k^2)
=-\zeta G_8
\frac\kappa3
\frac{A_0(M_\pi^2)}{(M_K^2+M_\pi^2)}\bigg[\frac{r^2-M_\pi^2}{r^2-M_K^2}\bigg]\\
+(M_\pi\leftrightarrow M_K, P\leftrightarrow r)\,,
\end{multline}
with the ``switch'' $\zeta=2\mp1$ for the ``exponential''~\eqref{eq:param_exp} and ``square-root''~\eqref{eq:param_sqrt} parametrizations, respectively.
For the neutral channel, this part is trivial:
Besides the fact that one does not need the off-shell version of the one-photon transition as the building block for the (absent) 1PR diagrams of \cref{fig:K->pigg}, the 1PR diagrams of \cref{fig:K->pig_LO_ChPT_a} do not arise either, and the 1PI part does not develop a nonvanishing off-shell part at LO in ChPT.
Regarding the $K^+\to\pi^+\gamma^*\gamma^*$ transition, it then follows that the final result at LO in ChPT presented in the next section does not receive any contribution from $\widetilde F(q^2)$ either; see also \cref{sec:gen_res_NLO}.

The (standard) $K\to\pi\gamma^*$ transition form factors $F(q^2)$ are physical (and unique) at LO in ChPT.
After the renormalization procedure, in which the UV-divergent parts stemming from the pion and kaon loops are absorbed by CTs, the results can be written as follows.
In the charged case of $K^+$,
\begin{equation}
\frac1{G_8}F^{(+)}(q^2)
=-\hat w_+
+F_\pi(q^2)+F_K(q^2)\,,
\label{eq:Fq2LO}
\end{equation}
with $\hat w_+=\frac23(12L_9^\text{r}-w_1^\text{r}-2w_2^\text{r})-\frac13(L_\pi+L_K)$, and for the neutral $K^0$ channel, the result is
\begin{equation}
\frac{\sqrt{2}}{G_8}F^{(0)}(q^2)
=\hat w_\text{S}
-2F_K(q^2)\,,
\label{eq:Fq2LO;0}
\end{equation}
with $\hat w_\text{S}=-\frac23(w_1^\text{r}-w_2^\text{r}+3w_3^\text{r})-\frac23L_K$.
Above,
\begin{equation}
F_P(q^2)
=\frac\kappa3\bigg[\frac13+\biggl(\frac{4M_P^2}{q^2}-1\biggr)\Jbar_P\bigl(q^2\bigr)\bigg]\,,
\label{eq:FPq2}
\end{equation}
with (by construction) finite $\Jbar_P$ defined in \cref{eq:Jbar}, and $\hat w_+$ and $\hat w_\text{S}$ are constants independent of the renormalization scale $\mu$.
Note that some simplifications can be achieved assuming the relation $w_2=4L_9$ and letting $w_3=0$~\cite{Ecker:1987qi}, leading to $\hat w_\text{S}=\hat w_++\frac13\kappa\log\frac{M_\pi^2}{M_K^2}$.
The listed results are related to those in Ref.~\cite{Ecker:1987qi} using $\hat w_{+,\text{S}}=2\kappa w_{+,\text{S}}$.
For completeness, the coefficients of the CT UV-divergent parts are found to be $\Gamma(L_9)=\frac14$, $\Gamma(w_1)=-1$.

\section{Two-photon transition}
\label{sec:2g}

In this section, I start by treating the charged-channel amplitude; the neutral channel is a special case of that.

\begin{figure}[t]
\centering
\includegraphics[width=0.98\columnwidth]{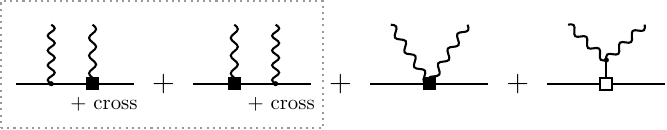}
\caption{
    The contributions to the $K\to\pi\gamma^*\gamma^*$ transition.
    The first two diagrams (within the gray box) are derived from the \mbox{$K^+\to\pi^+\gamma^*$} transition, which is denoted by three-particle vertices marked as filled black squares, and are absent in the neutral channel (they have photons attached to external legs).
    These relate to topology (1) of \cref{sec:2g}, and at LO in ChPT, the contributing diagrams are resolved by inserting those shown in \cref{fig:K->pig_LO_ChPT_a}.
    The third diagram is the remainder, which, resolved in LO ChPT, is shown in \cref{fig:K->pigg_LO_ChPT} and represents topology (2a), the ``genuine'' two-photon transition.
    The last diagram represents the contribution of the QCD anomaly and relates to topology (2b).
}
\label{fig:K->pigg}
\end{figure}

The (charged) two-photon-transition amplitude has three basic contributing topologies (see \cref{fig:K->pigg}):
\begin{enumerate}
    \item[(1)] The one induced by the one-photon transition (serving as a principal building block) with added ``bremsstrahlung'' photon radiated from (charged) meson legs (cf.\ the first two diagrams in \cref{fig:K->pigg}).
    \item[(2a)] The ``genuine'' two-photon contribution defined as the remainder of the full (nonanomalous) amplitude after the ``bremsstrahlung'' part [topology (1)] is subtracted (this is clarified later in terms of gauge invariance) (cf.~the third diagram in \cref{fig:K->pigg}).
    \item[(2b)] The QCD anomaly contribution, induced dominantly by $K\to\pi\pi^*$ and $K\to\pi\eta^*$ intermediate states (cf.~the last diagram in \cref{fig:K->pigg}).
\end{enumerate}
%
%
\begin{figure}[t]
\centering
\includegraphics[width=0.98\columnwidth]{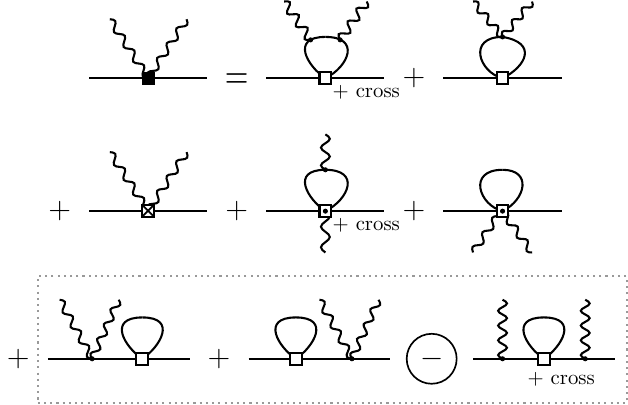}
\caption{
    The ``genuine'' two-photon contribution to the $K\to\pi\gamma^*\gamma^*$ transition.
    In the context of the decomposition in \cref{fig:K->pigg}, it is the remainder when the LO ChPT or its considered extensions are taken as a particular representation of \cref{fig:K->pigg}:
    Plugging the diagrams defined in \cref{fig:K->pig_LO_ChPT_a,fig:K->pigg_LO_ChPT} into \cref{fig:K->pigg} gives all the diagrams at LO ChPT.
    The minimal electromagnetic coupling is denoted by a small dot and the weak coupling \big($K\pi\pi\pi$ or $K\pi KK$\big) by an empty square; the combination of both is naturally represented by a square with a dot.
    Consequently, both (charged) pions and kaons are in the loops; the neutral-meson loops do not contribute at considered orders.
    The counterterm contributions are represented by a square with a cross.
    The ``+ cross'' note denotes that an associated diagram in which the photons are interchanged (crossed), i.e., Bose symmetrization, should be added.
    Notice the minus sign as the associated contribution needs to be subtracted due to double-counting in \cref{fig:K->pigg}.
    The neutral channel does not feature the three (plus one crossed) 1PR diagrams (with photons attached to external legs) placed within the gray dotted box.
    The LO counterterm contribution is also absent in the neutral channel.
}
\label{fig:K->pigg_LO_ChPT}
\end{figure}
%
All the contributing diagrams arising at LO in ChPT can then be identified from \cref{fig:K->pigg}, using the representation of the filled-black-square vertices, resolved by employing the graphical equations from \cref{fig:K->pig_LO_ChPT_a,fig:K->pigg_LO_ChPT}.%
\footnote{
    In this context, they are meant as a direct substitution rather than corresponding to matrix elements of another EFT Lagrangian.
}
Their sum leads to the physical $K\to\pi\gamma^*\gamma^*$ amplitudes --- gauge-invariant, UV-finite, and independent of the used parametrization.

Regarding the topology (1), a general discussion is needed before getting specifically to the LO ChPT case.
For instance, this contribution could be conveniently derived from a phenomenological form factor $F(q^2)$.
However, such a contribution (in general based on the first two diagrams of \cref{fig:K->pigg}) cannot stand alone, as this part of the amplitude itself would not be gauge invariant:
Additional contributions implicit in the third (contact) diagram of \cref{fig:K->pigg} are needed to obtain a consistent physical result.
The contribution of the first two (1PR) diagrams in \cref{fig:K->pigg} can be made gauge invariant by adding a minimal term in the spirit of Low's theorem~\cite{Low:1958sn} or using EFT techniques (with room for additional gauge-invariant terms; see \cref{sec:EFT}).
Such a minimally extended amplitude is proportional to
\begin{equation}
F_{\rho\sigma}\bigl(k_1,k_2\bigr)
=\widetilde F_{\rho\sigma}\bigl(k_1;k_2\bigr)
+\widetilde F_{\sigma\rho}\bigl(k_2;k_1\bigr)\,,
\label{eq:F_rs}
\end{equation}
with
\begin{multline}
\widetilde F_{\rho\sigma}\bigl(k_1;k_2\bigr)
=F(k_1^2)\\
\times\bigg\{
(k_1^2r_\rho-r\cdot k_1 k_{1\rho})\,\frac{(2P-k_2)_\sigma}{2P\cdot k_2-k_2^2}
+\big(P\leftrightarrow -r\big)\\
+\big(k_1^2g_{\rho\sigma}-k_{1\rho}k_{1\sigma}\big)
\bigg\}\,.
\label{eq:Ftilde_rs}
\end{multline}
This form satisfies $k_1^\rho\widetilde F_{\rho\sigma}(k_1;k_2)=k_2^\sigma\widetilde F_{\rho\sigma}(k_1;k_2)=0$; see also \cref{eq:Ftilde_rs_T4}.
Notice that $\widetilde F_{\rho\sigma}\bigl(k_1;k_2\bigr)$ vanishes, if $k_1$ is on-shell, and $\widetilde F_{\sigma\rho}\bigl(k_2;k_1\bigr)$ if $k_2$ is on-shell;
for both photons on-shell, the entire $F_{\rho\sigma}\bigl(k_1,k_2\bigr)$ vanishes, and consequently, the 1PR part does not contribute.

At LO in ChPT, the topology (1) is resolved by plugging the sum of diagrams in \cref{fig:K->pig_LO_ChPT_a} into the corresponding (i.e., the first two) diagrams in \cref{fig:K->pigg}.
To complete its gauge-invariant Lorentz structure represented by \cref{eq:F_rs,eq:Ftilde_rs} --- now specifically employing $F(q^2)$ of \cref{eq:Fq2LO} --- the necessary piece was hidden in the diagrams of \cref{fig:K->pigg_LO_ChPT}, the remaining (gauge-invariant) contribution of which is accounted for as the genuine two-photon transition (2a).
Adding (symbolically) the (anomalous) meson-pole part, the structure of the $K\to\pi\gamma^*\gamma^*$ amplitudes at LO is then
\begin{equation}
\begin{aligned}
&\mathcal{M}_{\rho\sigma}\big(K(P)\to\pi(r)\gamma_\rho^*(k_1)\gamma_\sigma^*(k_2)\big)\\
&=\xi
\Big[e^2F_{\rho\sigma}\bigl(k_1,k_2\bigr)
-2e^2G_8M_K^2\kappa\hat c\,
\hat T_{\rho\sigma}^{(1)}\Big]\\
&+\frac{\alpha}{2\pi}
G_8M_K^2
\frac{M_K^2}{2k_1\cdot k_2}
\Big[
A_\xi^{(1)}\hat T_{\rho\sigma}^{(1)}+A_\xi^{(2)}\hat T_{\rho\sigma}^{(2)}
\Big]\\
&+C_1T_{\rho\sigma}^{(a)}
\,,
\end{aligned}
\label{eq:Kpigg_LO}
\end{equation}
with $\xi=+1$ for the charged ($K^+$) and $\xi=0$ for the neutral ($K^0$) channels, and
\begin{equation}
\kappa\hat c
=\frac23(12L_9-w_1-2w_2)+\frac23\big(12L_{10}-2w_4\big)\,.
\label{eq:c_hat}
\end{equation}
Above, I introduced $2+1$ independent dimensionless gauge-invariant tensor structures
\begin{equation}
\begin{aligned}
\hat T_{\rho\sigma}^{(1)}&=\frac1{M_K^2}\big[(k_1\cdot k_2)g_{\rho\sigma}-{k_{1\sigma}k_{2\rho}}\big]\,,\\
\hat T_{\rho\sigma}^{(2)}
&=\frac4{M_K^2\lambda_s}\\
&\times\big[(k_1\cdot k_2)k_{1\rho}-k_1^2k_{2\rho}\big]
\big[(k_1\cdot k_2)k_{2\sigma}-k_2^2k_{1\sigma}\big]
\,,\\
T_{\rho\sigma}^{(a)}
&=\frac1{M_K^2}\,\varepsilon_{\rho\sigma\alpha\beta}k_1^\alpha k_2^\beta\,;
\end{aligned}
\label{eq:structs}
\end{equation}
notice that $\hat T_{\rho\sigma}^{(2)}$ vanishes if at least one of the photons is on-shell.%
\footnote{
    Meanwhile, it reduces to
    \begin{equation*}
        \hat T_{\rho\sigma}^{(2)**}
        =\frac{4k_1^2k_2^2}{M_K^2\lambda_s}\,{k_{1\sigma}k_{2\rho}}\,,
    \end{equation*}
    if both off-shell photons couple to conserved currents (as is the case, for instance, in $K\to\pi\ell_1^+\ell_1^-\ell_2^+\ell_2^-$ decays).}
The first term on the right-hand side of \cref{eq:Kpigg_LO} corresponds to the (stand-alone) contribution of topology (1), and the remainder, corresponding to topology (2a), is represented by the third line of \cref{eq:Kpigg_LO}, which by construction is also physical.
The last term relates to (2b), which this work does not discuss in detail; a relevant expression for $C_1$ can be found, e.g., in Eq.~(11) of Ref.~\cite{DAmbrosio:1996cak}.

The subamplitudes $A_\xi^{(i)}\equiv A_\xi^{(i)}(s,k_1^2,k_2^2)$, $i=1,2$, can be further decomposed based on the channel [charged (+) or neutral (0)] as
\begin{alignat}{3}
A_+^{(i)}
&={}&&\big[1-r_\pi^2+z\big]&&\hat A^{(i)}\big(M_\pi^2\big)\notag\\
&-{}&&\big[1-r_\pi^2-z\big]&&\hat A^{(i)}\big(M_K^2\big)\,,\label{eq:Ai+}\\
\frac1{\sqrt{2}}A_0^{(i)}
&={}&&\hphantom{1+{}}\big[r_\pi^2-z\big]&&\hat A^{(i)}\big(M_\pi^2\big)\notag\\
&-{}&&\big[1+r_\pi^2-z\big]&&\hat A^{(i)}\big(M_K^2\big)\,,\label{eq:Ai0}
\end{alignat}
with, finally, the two universal subamplitudes $\hat A^{(i)}\big(M_P^2\big)\equiv\hat A^{(i)}\big(M_P^2;s,k_1^2,k_2^2\big)$ that can be written in a very compact form:
\begin{align}
\hat A^{(1)}\big(M_P^2\big)
&=1+\frac2{\lambda_s}\Big\{
\big[\lambda_s{M_P^2}+k_1^2k_2^2s\big]C_0\big(s,k_1^2,k_2^2;M_P^2\big)\notag\\
&+k_1^2t_2\Jtilde_P\bigl(s,k_1^2\big)
+k_2^2t_1\Jtilde_P\bigl(s,k_2^2\big)
\Big\}\,,
\label{eq:A1_LO}
\end{align}
\begin{align}
&\hat A^{(2)}\big(M_P^2\big)
=-\hat A^{(1)}\big(M_P^2\big)
-\frac2{\lambda_s}(s-k_1^2-k_2^2)\notag\\
&\times\Big\{
t_1t_2C_0\big(s,k_1^2,k_2^2;M_P^2\big)
+t_1\Jtilde_P\bigl(s,k_1^2\big)
+t_2\Jtilde_P\big(s,k_2^2\big)
\Big\}\,.
\label{eq:A2_LO}
\end{align}
Above, $\lambda_s\equiv\lambda\big(s,k_1^2,k_2^2\big)=4(k_1\cdot k_2)^2-4k_1^2k_2^2$, with $\lambda$ being the triangle K\"all\'en function, $t_{1,2}=k_{1,2}\cdot(k_1+k_2)$, $s=t_1+t_2=(k_1+k_2)^2$, $z=s/M_K^2$, $r_\pi=M_\pi/M_K$, and the loop functions are defined in \cref{sec:functions}.
To my knowledge, this result has not been derived before.

Although the results for special cases with one or two on-shell photons can be readily read off, I state them below for completeness.
They can be written as
\begin{align}
\begin{split}
&\mathcal{M}_{\rho\sigma}\big(K(P)\to\pi(r)\gamma_\rho(k_1)\gamma_\sigma(k_2)\big)\\
&\quad=\frac\alpha{2\pi}G_8M_K^2
\biggl[-\xi\hat c+\frac1{2z_0}A_\xi^{(1)}(s,0,0)\bigg]\hat T_{\rho\sigma}^{(1)}
+C_1T_{\rho\sigma}^{(a)}\,,
\label{eq:Kpigg_LO_gg}
\end{split}\\
\begin{split}
&\mathcal{M}_{\rho\sigma}\big(K(P)\to\pi(r)\gamma_\rho^*(k_1)\gamma_\sigma(k_2)\big)\\
&\quad=\xi e^2\widetilde F_{\rho\sigma}\bigl(k_1;k_2\bigr)\Big|
_{\substack{\scalebox{0.8}{$\displaystyle k_2^2\,\to\,0,\,k_{2\sigma}[\epsilon^\sigma(k_2)]\to\,0$}}}\\
&\quad+\frac\alpha{2\pi}G_8M_K^2
\biggl[-\xi\hat c+\frac1{2z_0}A_\xi^{(1)}(s,k_1^2,0)\bigg]\hat T_{\rho\sigma}^{(1)}
+C_1T_{\rho\sigma}^{(a)}\,,
\label{eq:Kpigg_LO_g*g}
\end{split}
\end{align}
with $z_0=k_1\cdot k_2/M_K^2$.
The doubly off-shell subamplitude $\hat A^{(1)}$ from \cref{eq:A1_LO} and entering \cref{eq:Ai+,eq:Ai0} is thus reduced to
\begin{align}
\hat A^{(1)}\big(M_P^2;s,0,0\big)
&=1+2M_P^2C_0\big(s,0,0;M_P^2\big)\,,\label{eq:Agg(1)}\\
\begin{split}
\hat A^{(1)}\big(M_P^2;s,k_1^2,0\big)
&=1+2M_P^2C_0\big(s,k_1^2,0;M_P^2\big)\\
&+\frac{k_1^2\Jtilde_P\big(s,k_1^2\big)}{s-k_1^2}\,,
\end{split}
\end{align}
for the doubly on-shell [\cref{eq:Kpigg_LO_gg}] and semi-off-shell [\cref{eq:Kpigg_LO_g*g}] cases, respectively.
The relation of these compact expressions with the literature is based on relating the $\bar J$ and $C_0$ functions with $F(z)$ and $R(z)$ of Refs.~\cite{Ecker:1987fm,Cohen:1993ta}; see the end of \cref{sec:functions}.

\section{The generic ansatz}
\label{sec:general}

The structure of \cref{eq:Ai+,eq:Ai0} motivates calculating the same set of diagrams as in the LO ChPT case but with a generic $K\to\pi PP$ vertex to simultaneously account for all the possible channels (and potentially other related processes).
The following subsections introduce such a vertex as polynomials at $\mathcal{O}(p^2)$ and $\mathcal{O}(p^4)$.

\begin{figure}[t]
\centering
\includegraphics[width=0.98\columnwidth]{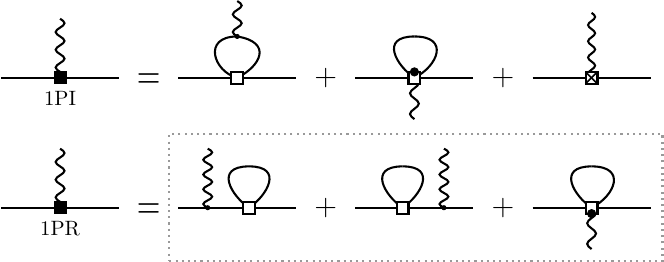}
\caption{
    The representation of the $K\to\pi\gamma^*$ transition amplitude in terms of LO Feynman diagrams emerging from the ChPT Lagrangian~\eqref{eq:L_ChPT}, while utilizing the diagrammatic notation of \cref{sec:notation}.
    The caption of \cref{fig:K->pig_LO_ChPT_a} applies.
    Furthermore, the one-particle-irreducible (1PI) (radiative) tadpole contribution is separated in two pieces (see also \cref{sec:notation}) for the diagrams in respective lines to form gauge-invariant subsets.
    Strictly speaking, the labeling 1PI and 1PR is thus violated; here, the `R' in the latter can be thought of as standing for ``reducible and remainder.''
}
\label{fig:K->pig_LO_ChPT_b}
\end{figure}

\begin{figure}[t]
\centering
\includegraphics[width=0.84\columnwidth]{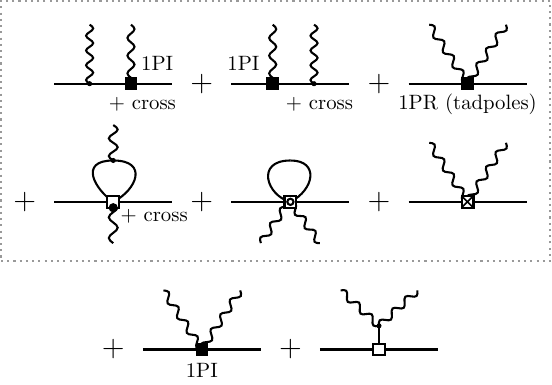}
\caption{
    The relevant contributions to the $K\to\pi\gamma^*\gamma^*$ transition at LO ChPT and its considered extensions.
    The first two diagrams within the gray box (present only in the charged channel) are derived from the \mbox{$K^+\to\pi^+\gamma^*$} transition, which is denoted by vertices marked as filled black squares:
    Explicitly, the contributing diagrams are recovered by plugging in the 1PI vertices from \cref{fig:K->pig_LO_ChPT_b}, and to form a gauge-invariant subset, they are accompanied by the related diagrams (shown in the second line) where a photon is radiated directly from the vertices of 1PI diagrams of \cref{fig:K->pig_LO_ChPT_b} (associated with external meson radiation in the sense of \cref{sec:notation}).
    The genuine two-photon transition (structure-dependent part) is the 1PI remainder, which, resolved in LO ChPT and considered extensions, is shown in \cref{fig:K->pigg_LO_1PI}.
    The remaining tadpole diagrams are in \cref{fig:K->pigg_LO_1PR}.
    The diagrams within the gray dotted box are absent in the neutral channel.
}
\label{fig:K->pigg_LO}
\end{figure}

\begin{figure}[th]
\centering
\includegraphics[width=0.84\columnwidth]{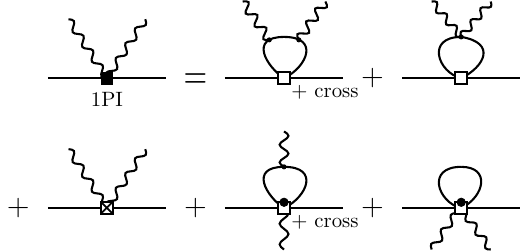}
\caption{
    The ``genuine'' two-photon contribution to the $K\to\pi\gamma^*\gamma^*$ transition in the diagrammatic notation of \cref{sec:notation}.
    In the context of the decomposition in \cref{fig:K->pigg}, it is the remainder when the LO ChPT or its considered extensions are taken as a particular representation of \cref{fig:K->pigg}.
    As defined here, these diagrams form a gauge-invariant subset.
}
\label{fig:K->pigg_LO_1PI}
\end{figure}

\begin{figure}[th]
\centering
\includegraphics[width=0.98\columnwidth]{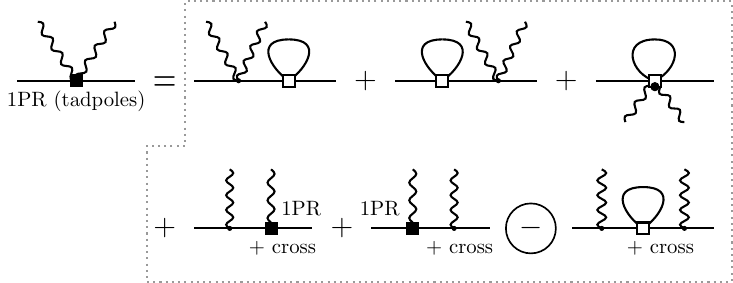}
\caption{
    Remaining tadpole diagrams, the sum of which vanishes at considered orders.
    Notice the minus sign as the associated contribution needs to be subtracted due to double-counting in the sum of the two preceding diagrams.
    These diagrams are all in the gray box as they are absent in the neutral channel.
}
\label{fig:K->pigg_LO_1PR}
\end{figure}

Moreover, one can notice that the standard diagrams used in the previous sections, in particular in \cref{fig:K->pig_LO_ChPT_a,fig:K->pigg_LO_ChPT}, do not respect some symmetries of the amplitude:
In particular, they do not reflect Ward identities for the respective underlying currents.
Consequently, a gauge-invariant result emerges only after summing all diagrams.
Instead, it is possible and beneficial to group the contributing diagrams so that they form gauge-invariant subsets, some of which potentially vanish --- and it happens to be the case.
This is achieved by separating the radiative vertices $K\pi PP\gamma(\gamma)$ by pairing the charged fields in a suitable way:
Specifically, I separate the cases in which internal (loop) and external mesons are treated as charged.
The alternative set of diagrams (\cref{fig:K->pig_LO_ChPT_b,fig:K->pigg_LO,fig:K->pigg_LO_1PI,fig:K->pigg_LO_1PR}) is then introduced and used instead; see \cref{sec:notation} for details.

As an immediate example, the alternative set of diagrams for the one-photon transition can be seen in \cref{fig:K->pig_LO_ChPT_b}.
Everything stays the same except for the radiative vertex, which is separated into two contributions.
This allows the 1PI part to be gauge-invariant and for the 1PR part to vanish on-shell (or identically for some set of vertex-ansatz coefficients), relating thus closely the charged and neutral channels (in the latter, there is no 1PR contribution).

%
%
%

\subsection{Leading-order analysis}

For a general off-shell four-particle vertex at $\mathcal{O}(p^2)$, from four momenta, six scalar products and four virtualities can be built.
From the 4-momentum conservation at the vertex, three plus one of these scalar products can be written as linear combinations of the rest.
After removing these, a general LO $K\pi P\bar P$-vertex amplitude can be written as
\begin{equation}
\begin{aligned}
\frac1{\GF}&\mathcal{M}^{(2)}\bigl(K(P)\pi(r)P(p_1)\bar P(p_2)\bigr)\\
&=aM_K^2+b_0p_1\cdot p_2+b_+(P\cdot p_1+r\cdot p_2)\\
&\qquad+a_K\bigl(P^2-M_K^2\bigr)+a_\pi\bigl(r^2-M_\pi^2\bigr)\\
&\qquad+a_1\bigl(p_1^2-M_P^2\bigr)+a_2\bigl(p_2^2-M_P^2\bigr)\,,
\label{eq:KpiPP_LO_gen}
\end{aligned}
\end{equation}
where all momenta are incoming.
All the physical cases (including the channels discussed in previous sections) can be brought to the form in \cref{eq:KpiPP_LO_gen}; see \cref{sec:amplitudes} for the LO $K\pi P\bar P$ amplitudes in ChPT.

I will employ the ansatz \eqref{eq:KpiPP_LO_gen} to calculate the generic LO amplitude for the $K\to\pi\gamma^{(*)}[\gamma^{(*)}]$ processes.
This approach allows for the simultaneous evaluation of all considered channels, observation of their relations, and discussion of some general properties.
It further uncovers what parameters the final physical amplitude depends on and which parts of the amplitude are irrelevant (including the diagrams in \cref{fig:K->pigg_LO_1PR}).
Finally, it sets the stage for going beyond the LO calculation.

As an example, consider the charged channel and the transition amplitudes $K^+\to\pi^+\gamma^{(*)}[\gamma^{(*)}]$.
The main ingredient for the LO ChPT calculation is the LO \mbox{$K^+\to\pi^+\pi^+\pi^-$} amplitude.
Lagrangian~\eqref{eq:L_ChPT} gives, in different parametrizations, terms shown in \cref{sec:Lagrangians}, which lead to the on-shell amplitudes shown in \cref{sec:amplitudes}.
In particular, one arrives at
\begin{equation}
\frac1{G_8}\mathcal{M}^{(2)}\bigl(K^+(P)\to\pi^+\pi^-(p_1)\pi^+\bigr)
=-2P\cdot p_1\,.
\end{equation}
Employed parametrizations differ in their off-shell terms [i.e., the coefficients $a_K$, $a_\pi$, $a_1$, and $a_2$ of \cref{eq:KpiPP_LO_gen}, which vanish when mesons are on-shell], although they yield the same on-shell amplitudes.
To be specific, the exponential parametrization~\eqref{eq:param_exp} gives, in addition to values in \cref{tab:amplitudes} and in the units of $\go$,
$a_K^\text{(exp)}=-\frac12$, $a_\pi^\text{(exp)}=\frac16$, $a_1^\text{(exp)}=\frac56$, and $a_2^\text{(exp)}=\frac16$,
while the ``square-root'' parametrization~\eqref{eq:param_sqrt} gives
$a_K^\text{(sqrt)}=a_\pi^\text{(sqrt)}=a_1^\text{(sqrt)}=a_2^\text{(sqrt)}=-\frac12$.
Similarly, one can write
\begin{equation}
\frac1{G_8}\mathcal{M}^{(2)}\bigl(K^+K^+\to\pi^+(r)K^+(p_2)\bigr)
=2r\cdot p_2\,,
\end{equation}
plus terms vanishing on-shell.
In particular, after the on-shell part is fixed to have the form of the ansatz~\eqref{eq:KpiPP_LO_gen} with the values from \cref{tab:amplitudes}, one finds
$a_K^\text{(exp)}=\frac16$, $a_\pi^\text{(exp)}=-\frac12$, $a_1^\text{(exp)}=\frac16$, $a_2^\text{(exp)}=\frac56$,
and again
$a_K^\text{(sqrt)}=a_\pi^\text{(sqrt)}=a_1^\text{(sqrt)}=a_2^\text{(sqrt)}=-\frac12$.

Serving as a consistency check, these off-shell terms do not affect the resulting physical amplitudes for $K^+\to\pi^+\gamma^{(*)}[\gamma^{(*)}]$ transitions.
On the other hand, with an appropriate (off-shell-scheme) choice of $a_I$, one can restrict the number of diagrams necessary to evaluate.
Indeed, one can remove the $K^+\to\pi^+\pi^+\pi^-\gamma^{(*)}$ and $K^+\to\pi^+\pi^+\pi^-\gamma^{(*)}\gamma^{(*)}$ vertices (stemming from the effective Lagrangian leading to the correct on-shell $K^+\to\pi^+\pi^+\pi^-$ amplitude) and thus reduce the number of cancellations among the diagrams necessary to lead to gauge-invariant $K^+\to\pi^+\gamma^{(*)}[\gamma^{(*)}]$ amplitudes;
in LO ChPT, this occurs naturally by employing parametrization \cref{eq:param_sqrt}.
Another, probably a more relevant choice, leads to the fact that some other (tadpole) diagrams evaluate to zero.
This point is explained in detail in \cref{sec:tadpole_gen}.
Nevertheless, working with a generic parametrization is still a very rewarding option, as employing the diagrammatic notation of \cref{sec:notation} already results in many simplifications.

\subsection{Unitarity corrections from $K\to3\pi$}
\label{sec:NLO}

The dominant NLO corrections to the $K\to\pi\gamma^*$~\cite{DAmbrosio:1998gur}, $K\to\pi\gamma\gamma$~\cite{Cappiello:1992kk,Cohen:1993ta,DAmbrosio:1996cak}, and $K\to\pi\gamma\gamma^*$~\cite{Donoghue:1997rr,Donoghue:1998ur,Gabbiani:1998tj} transitions were evaluated in terms of the so-called unitarity corrections in the given references.
The idea is to take the LO ChPT calculation and related Feynman diagrams, as described in the previous sections, and replace the LO \big[$\mathcal{O}\big(p^2\big)$\big] $K\pi PP$ vertex with a phenomenological $\mathcal{O}\big(p^4\big)$ amplitude consistent with the $K\to\pi\pi^+\pi^-$ ($K\to3\pi$) data.
Such an (EFT-Lagrangian-driven) approach is equivalent to dispersive methods, and the role of a subtraction polynomial is replaced by a polynomial with free parameters stemming from appropriate CTs.
Here, I present the equivalent calculation for the doubly off-shell case of $K\to\pi\gamma^*\gamma^*$;
the previously obtained results are special cases of that.

The $K\to3\pi$ decay data and the quadratic fits of the associated Dalitz plots lead to $K\to3\pi$ amplitudes, typically written in a form such as~\cite{Cappiello:1992kk,DAmbrosio:1996cak,DAmbrosio:1998gur,Gabbiani:1998tj}
\begin{align}
    &\mathcal{M}\bigl(K^+(P)\to\pi^+(q_1)\pi^-(q_3)\pi^+(q_2)\bigr)\notag\\
    &=\alpha_++\beta_+ Y+\gamma_{1,+}\big(Y^2+\tfrac13X^2\big)+\gamma_{2,+}\big(Y^2-\tfrac13X^2\bigr)\,,\\
    &\mathcal{M}\bigl(\KL(P)\to\pi^0(q_3)\pi^-(q_1)\pi^+(q_2)\bigr)\notag\\
    &=\alpha_\text{L}+\beta_\text{L} Y+\gamma_{1,\text{L}}\big(Y^2+\tfrac13X^2\big)+\gamma_{2,\text{L}}\big(Y^2-\tfrac13X^2\bigr)\,,\\
    &\mathcal{M}\bigl(\KS(P)\to\pi^0(q_3)\pi^-(q_1)\pi^+(q_2)\bigr)\notag\\
    &=\beta_\text{S}X+\gamma_\text{S} XY\,,
\end{align}
with $X=\frac1{M_\pi^2}(s_2-s_1)$, $Y=\frac1{M_\pi^2}(s_3-s_0)$, $s_0=\frac13\sum_{i=1}^3s_i$, and $s_i=(P-q_i)^2$.
Assuming real coefficients, squares of the above $K^+$ and $\KL$ decay amplitudes to be compared with the Dalitz plot fits~\cite{ParticleDataGroup:2024cfk} are
\begin{equation}
\begin{aligned}
    |\mathcal{M}_\xi|^2
    &=\alpha_\xi^2\big[1+2\hat\beta_\xi Y\\
    &\hspace{1cm}+\big(2\hat\gamma_{1,\xi}+\tfrac12\hat\beta_\xi^2\big)\big(Y^2+\tfrac13X^2\big)\\
    &\hspace{1cm}+\big(2\hat\gamma_{2,\xi}+\tfrac12\hat\beta_\xi^2\big)\big(Y^2-\tfrac13X^2\big)\big]\,,
\end{aligned}
\end{equation}
with $\xi=+$ or $\text{L}$, $\hat\beta_\xi=\beta_\xi/\alpha_\xi$, and $\hat\gamma_{i,\xi}=\gamma_{i,\xi}/\alpha_\xi$.

Irrespective of the particular form or parametrization, regarding on-shell amplitudes, the terms quartic in momenta can always be brought to the form
\begin{multline}
\frac1{\GF}\mathcal{M}^{(4)}\bigl(K(P)\pi(r)P^+(p_1)P^-(p_2)\bigr)\\
=\sum_{\pm}\cc_\pm\frac1{\ChPT^2}\big[(P\cdot p_1)(p_2\cdot r)\pm(P\cdot p_2)(p_1\cdot r)\big]\\[-2mm]
+\cc_3\frac1{\ChPT^2}(P\cdot r)(p_1\cdot p_2)\,,
\label{eq:K3pi_NLO_gen}
\end{multline}
with a remainder that can be written in the form of LO ansatz~\eqref{eq:KpiPP_LO_gen};
I use $\ChPT=4\pi F_\pi$ for the expansion scale.
In other words, the higher-order pion-loop-induced unitarity corrections to the general LO result can be represented by the terms proportional to the additional three coefficients $c_i$.
Similarly to the $\mathcal{O}(p^2)$ vertex amplitude \eqref{eq:KpiPP_LO_gen}, all possible off-shell extensions could be added to \cref{eq:K3pi_NLO_gen}.
For consistency, I explicitly checked that such $\mathcal{O}(p^4)$ off-shell terms do not affect the final results for the on-shell amplitudes either, and I do not include them in \cref{eq:K3pi_NLO_gen} for brevity.
This follows from the corresponding Lagrangian utilizing equations of motion.

\section{General results at LO and beyond}
\label{sec:gen_res_NLO}

In this section, I consider, for the $K\pi PP$ vertex amplitude, the ansatz given by the sum of \cref{eq:KpiPP_LO_gen,eq:K3pi_NLO_gen}, i.e.,
\begin{multline}
 \mathcal{M}_{K\pi PP}^\text{(2+4)}\bigl(K\pi P\bar P\bigr)\\
 \equiv\mathcal{M}_\eqref{eq:KpiPP_LO_gen}^{(2)}\bigl(K\pi P\bar P\bigr)
 +\mathcal{M}_\eqref{eq:K3pi_NLO_gen}^{(4)}\bigl(K\pi P\bar P\bigr)\,,
\label{eq:KpiPP_NLO_sum}
\end{multline}
where I have suppressed the explicit particle momenta.
The diagrams are the same as in the previous sections, but the LO ChPT input is replaced by a more general case~\eqref{eq:KpiPP_NLO_sum}.

Before the diagrams can be evaluated, these four-meson vertices need to be supplemented with their radiative counterparts.
A standard approach is to write an effective Lagrangian that leads to amplitude \eqref{eq:KpiPP_NLO_sum} and replace all partial derivatives with covariant ones.
First, note that it is beneficial to do this in stages (not all derivatives at once), which leads to an alternative diagrammatic notation of \cref{sec:notation}.
Second, as the amplitude~\eqref{eq:K3pi_NLO_gen} is linear in all four 4-momenta, one can directly work with the amplitude and modify it in the following way:
\begin{enumerate}
    \item[1)] Shift the momenta, $$p_\mu\to p_\mu+\eta_p e\epsilon_\mu\,,$$ where $\eta_p$ is the charge (in the units of $e$) of the incoming particle associated with momentum $p$.
    \item[2)] Replace the polarization vectors,
    \begin{enumerate}
        \item[a)] $e\epsilon_\alpha\to eg_{\alpha\rho}$\,,
        \item[b)] $e^2\epsilon_\alpha\epsilon_\beta\to e^2(g_{\alpha\rho}g_{\beta\sigma}+g_{\alpha\sigma}g_{\beta\rho})$\,,
    \end{enumerate}
    in terms proportional to $e$ and $e^2$, respectively.
\end{enumerate}
This can be done because, in the corresponding effective Lagrangian, no derivatives of photon fields occur when the covariant derivatives are introduced, as there is only one derivative acting on each field.
This is in contrast with the case when various off-shell extensions [as in \cref{eq:KpiPP_LO_gen}] are considered:
For terms that are proportional to the squares of momenta, e.g., $P^2$, after the above momentum shift is performed, one has to perform yet another step:
\begin{enumerate}
    \item[3)] Substitute further
    \begin{enumerate}
        \item[a)] $e^0p^2\to p^2+\eta_p ek_\rho$\,,
        \item[] and at higher orders,
        \item[b)] $ep^2\epsilon_\alpha\to ep^2\epsilon_\alpha+\eta_P e^2(k_{1\rho}g_{\alpha\sigma}+k_{2\sigma}g_{\alpha\rho})$\,,
        \item[c)] $e^0p_1^2p_2^2\to p_1^2p_2^2+2\eta_{p_1}\eta_{p_2} e^2k_{1\rho}k_{2\sigma}$\,.
    \end{enumerate}
\end{enumerate}
These additional terms vanish on-shell or after contracting with conserved currents or if a specific gauge is used.
Moreover, in the present application, they are only related to the off-shell extensions that do not affect the final physical expressions.
For completeness, the resulting amplitudes are listed in \cref{sec:notation}, employing a related technique.

\subsection{The tadpole}
\label{sec:tadpole_gen}

The main building block of 1PR diagrams presented in \cref{fig:K->pig_LO_ChPT_b,fig:K->pigg_LO_1PR} is the (tadpole) (mixing) transition
\begin{multline}
\frac1{\GF}\mathcal{M}\big(K(p)\to\pi(p)\big)
=-\kappa A_0(M_P^2)\Big\{g_P(p^2)\\
+\big[c_3+\tfrac12c_+\big]\frac{M_P^2}{\ChPT^2}p^2\Big\}-\frac\kappa4c_+\frac{M_P^4}{\ChPT^2}p^2
\,,
\label{eq:tadpole_gen}
\end{multline}
with
\begin{equation}
g_P(p^2)=aM_K^2-b_0M_P^2+a_K(p^2-M_K^2)+a_\pi(p^2-M_\pi^2)
\end{equation}
and $M_P$ the mass of the mesons running in the loop.
I now discuss several cases.

At $\mathcal{O}(p^4)$ [corresponding to the first line of \cref{eq:tadpole_gen} only, i.e., the terms proportional to coefficients $c_i$ are absent], $g_P(p^2)$ --- and consequently the whole amplitude $\mathcal{M}\big(K(p)\to\pi(p)\big)$ --- vanishes with the choice
\begin{equation}
    a_\pi=-a_K\,,\qquad
    a_K=\frac{a-b_0r_P^2}{1-r_\pi^2}\,,
\end{equation}
where $r_P=M_P/M_K$.
The corresponding (particular) field redefinition thus leads to significant simplifications.

At $\mathcal{O}(p^6)$, one does not seem to have such a comparably natural choice, as $(P\cdot p_{1,2})(p_{2,1}\cdot r)$ [as in \cref{eq:K3pi_NLO_gen}] are the only combinations of momenta appearing in the $K\pi PP$ vertex that lead to terms not proportional to $A_0(M^2)$.
On the other hand, one can formally choose
\begin{equation}
    a_\pi=-a_K-\tilde c\,,\qquad
    a_K=\frac{a-b_0r_P^2+\tilde c\,r_\pi^2}{1-r_\pi^2}\,,
\end{equation}
with
\begin{equation}
    \tilde c
    =\frac{M_P^2}{\ChPT^2}\bigg(c_3+\frac12c_++\frac14\frac{M_P^2}{A_0(M_P^2)}\,c_+\bigg)\,.
\end{equation}

However, irrespective of the parametrization choice, notice in \cref{eq:tadpole_gen} that, at considered orders, $f(p^2)\equiv\mathcal{M}\big(K(p)\to\pi(p)\big)$ is linear in $p^2$.
In the following, I consider $K$ and $\pi$ charged.
Adding a photon to each of the two (charged) external legs (as in the second line of \cref{fig:K->pig_LO_ChPT_b}) leads to the following amplitude:
\begin{equation}
    -e(P+r)_\rho
    \bigg[
    \frac{f(P^2)}{P^2-M_\pi^2}
    -\frac{f(r^2)}{M_K^2-r^2}
    \bigg]\,,
\label{eq:A0+g}
\end{equation}
which reduces for on-shell $K$ and $\pi$ to
\begin{equation}
    -e(P+r)_\rho f'(0)\,.
\end{equation}
And this is, up to the sign, the associated tadpole diagram with an extra vertex photon (in which only the {\em external} fields are considered charged and thus only the external momenta are shifted or taken derivative with respect to in the sense of \cref{sec:notation}), i.e., a loop integral of the matrix element in \cref{{eq:M_KpiPPg_E}}:
\begin{multline}
    \int\frac{\d^4\ell}{(2\pi)^4}\frac{i^2}{\ell^2-M_P^2+i\epsilon}\\
    \times\mathcal{M}_{\text{(E)},\,\rho}^\text{(2+4)}\bigl(K^+(P)\pi^-(-r)P(\ell)\bar P(-\ell)\gamma_\rho(r-P)\bigr)\,.
\end{multline}
In other words, a direct calculation shows that, up to $\mathcal{O}(p^6)$ (regarding the considered extension), the sum of the 1PR diagrams with their corresponding radiative-vertex tadpole (now the whole second line of \cref{fig:K->pig_LO_ChPT_b} that forms a gauge-invariant subset) vanishes once the mesons are on-shell.
This is related to the underlying Ward identity.
At the studied orders, the full one-photon transition amplitude is thus given solely by the 1PI diagrams.
Moreover, even when the contribution of the 1PR part to the {\em two-photon} transition is considered in the sense of the second line of \cref{fig:K->pigg_LO_1PR}, these do not contribute at considered orders either:
The diagrams in \cref{fig:K->pigg_LO_1PR} form a gauge-invariant subset and their sum vanishes on-shell.
This is connected to the fact that the tadpole itself is unphysical on-shell, and one can choose the off-shell continuation of the amplitude~\eqref{eq:tadpole_gen} (the associated arbitrary parameters $a_I$) so that it disappears, together with all the 1PR diagrams that contain it or the corresponding diagrams with extra photons derived from it.

\subsection{The one-photon transition}
\label{sec:1g_gen}

In this subsection, I present the result for the on-shell one-photon transition.
As justified in the previous subsection, at given orders, it is sufficient to consider only the 1PI diagrams of \cref{fig:K->pig_LO_ChPT_b}, which add up to a physical amplitude, i.e., gauge-invariant, UV-finite, and independent of off-shell extrapolations.

In general, the Lorentz symmetry and gauge invariance dictate
\begin{multline}
\mathcal{M}_\rho\big(K(P)\to\pi(r)\gamma_\rho^*(k)\big)\\
=eF(k^2)
\big[
k^2r_\rho
-(k\cdot r)k_\rho
\big]\,.
\label{eq:M_K->Pg}
\end{multline}
The ansatz~\eqref{eq:KpiPP_NLO_sum} leads to a generic form factor that can be written as
\begin{equation}
    \frac1\GF F(q^2)
    =F^\text{(CT)}(q^2)
    +\sum_P\widehat F_P(q^2)\,,
\end{equation}
with
\begin{align}
    F^\text{(CT)}(q^2)
        &=F_\text{UV-fin.}^\text{(CT)}(q^2)+
        F_\text{UV-div.}^\text{(CT)}(q^2)\,,\\
        \widehat F_P(q^2)
        &=h_P(q^2)F_P(q^2)
        +\widehat F_P^\text{(UV+log)}(q^2)\,,
\end{align}
where the remaining UV-divergent part included in
\begin{equation}
\widehat
F_P^\text{(UV+log)}(q^2)
=-\widehat h_P(q^2)\bigg(\frac\kappa2\frac1{\tilde\epsilon}-\frac12L_P\bigg)
\label{eq:F_P_UV}
\end{equation}
must be absorbed by the CT contribution $F^\text{(CT)}(q^2)$ [canceled against $F_\text{UV-div.}^\text{(CT)}(q^2)$],
\begin{align}
    h_P(q^2)
    &=b_+^{(P)}-\frac{c_-^{(P)}}2\frac{q^2}{\ChPT^2}\,,\label{eq:h(q^2)}\\
    \widehat h_P(q^2)
    &=\frac23h_P(q^2)+2c_-^{(P)}\frac{M_P^2}{\ChPT^2}\,,\label{eq:hhat(q^2)}
\end{align}
and where $F_P(q^2)$ is the same as in \cref{eq:FPq2}, now multiplied by a generic polynomial factor $h_P(q^2)$.
Combining the LO and NLO CTs discussed in \cref{sec:CTs} leads to the generic CT of three terms:
\begin{equation}
\begin{aligned}
F^\text{(CT)}(q^2)
&=-c^{(4)}
-\cA_1\frac{M_K^2}{\ChPT^2}+c^{(6)}_2\frac{q^2}{\ChPT^2}\,.
\end{aligned}
\end{equation}
It is easy to see that this polynomial is indeed capable of canceling the divergences arising in $\sum_P\widehat F_P(q^2)$, i.e., has the same form as the polynomial in \cref{eq:F_P_UV}.
The particular UV-divergences canceling is not relevant for this work, but I list it for completeness.
At LO, from \cref{eq:F_P_UV}, $\widehat F_P(q^2)\big|_\text{UV}\propto-\frac23b_+$, so letting $\Gamma\big(c^{(4)}\big)=\sum_P\frac23b_+^{(P)}$ leads to a convergent form factor.
At NLO, one then adds $\Gamma\big(\cA_1\big)=\sum_P2c_-^{(P)}\frac{M_P^2}{M_K^2}$, $\Gamma\big(c_2^{(6)}\big)=\sum_P\frac13c_-^{(P)}$.

It is thus clear that one can simply write
\begin{equation}
    \frac1\GF F(q^2)
    =F_\text{pol.}(q^2)+\sum_Ph_P(q^2)F_P(q^2)\,,
\label{eq:F_gen}
\end{equation}
with the polynomial
\begin{equation}
    F_\text{pol.}(q^2)
    =F_\text{UV-fin.}^\text{(CT)}(q^2;\mu)
    +\frac12\sum_P\widehat h_P(q^2)L_P(\mu)
\label{eq:F_pol}
\end{equation}
being channel specific and independent of $\mu$.

Out of all parameters in the ansantz \eqref{eq:KpiPP_NLO_sum}, the result \eqref{eq:F_gen} depends only on $b_+$ and $c_-$.
Restricting the result to LO ChPT, it is proportional, modulo the CT finite part, to $h_P(q^2)=b_+^{(P)}$ only.
As a direct consequence, for the charged channel, both the charged-pion and charged-kaon loops contribute as $b_+=\go$ in both cases, while for the neutral channel, only the charged-kaon loops contribute with $b_+=-\sqrt{2}\go$; see \cref{tab:amplitudes} in \cref{sec:amplitudes} for LO ChPT amplitudes.
This is consistent with the results in \cref{eq:Fq2LO,eq:Fq2LO;0} in \cref{sec:1g}.
Indeed, at LO in ChPT, the CT contribution $F^\text{(CT)}(q^2)=-c^{(4)}$, with $c^{(4)}$ shown explicitly in \cref{eq:c/d(4)+,eq:c/d(4)0}, absorbs the UV divergences, and the remaining finite part $F_\text{UV-fin.}^\text{(CT)}$ together with the logarithm part featuring $\widehat h_P(q^2)=\frac23b_+^{(P)}$ leads to $F_\text{pol.}^{(+)}(q^2)=-\hat w_+$ and $F_\text{pol.}^{(0)}(q^2)=\hat w_\text{S}$.

\subsection{The two-photon transition}
\label{sec:2g_gen}

As discussed in \cref{sec:tadpole_gen}, the tadpole contributions of \cref{fig:K->pigg_LO_1PR} do not need to be considered in the case following the ansatz \eqref{eq:KpiPP_NLO_sum}:
They form a gauge-invariant subset and vanish on-shell.
This reduces the number of relevant contributions in \cref{fig:K->pigg_LO}.
The 1PR part stemming from the 1PI diagrams of the one-photon transition (present in the charged channel) and supplemented with associated tadpole and CT contributions (cf.~the middle line of \cref{fig:K->pigg_LO}) is implemented in terms of \cref{eq:F_rs,eq:Ftilde_rs} [plugging in the form factor~\eqref{eq:F_gen}] and a gauge-invariant remainder.
As the resulting contribution is conveniently stand-alone (gauge invariant and finite), a phenomenological form factor $F(q^2)$ extracted from $K^+\to\pi^+\ell^+\ell^-$ data can also be used.

Employing the ansatz \eqref{eq:KpiPP_NLO_sum}, a direct computation of the 1PI part (explicitly shown in \cref{fig:K->pigg_LO_1PI}) leads to the origin of the factorization in \cref{eq:Ai+,eq:Ai0}.
Except for the NLO terms proportional to $c_+$, these expressions can be reduced to
\begin{equation}
A^{(i)}
=\sum_Pj_P(z)\hat A^{(i)}\big(M_P^2\big)\,,\qquad i=1,2\,,
\label{eq:A_ch}
\end{equation}
with
\begin{equation}
    j_P(z)
    =-2a^{(P)}+b_+^{(P)}z
    +\bigg[b_0^{(P)}-\tilde c_3^{(P)}\frac{P\cdot r}{M_K^2}\bigg]\big(2r_P^2-z\big)\,.
\label{eq:jP}
\end{equation}
Above, I used $\tilde c_3^{(P)}=c_3^{(P)}/r_\chi^2$, where $r_\chi=\frac{\ChPT}{M_K}$, and I kept $P\cdot r=\frac12M_K^2(1+r_\pi^2-z)$.
In particular, plugging the coefficients from \cref{tab:amplitudes} of \cref{sec:amplitudes} into \cref{eq:jP} restores \cref{eq:Ai+,eq:Ai0}.
Needless to say, the subamplitudes $\hat A^{(i)}\big(M_P^2\big)$ are the same as in \cref{eq:A1_LO,eq:A2_LO}.

In contrast with the one-photon transition described in \cref{sec:1g_gen}, the two-photon transition depends on many (on-shell relevant) parameters of the ansatz \eqref{eq:KpiPP_NLO_sum}, but $c_-$ is not one of them.
It was already shown in \cref{eq:jP} that the terms proportional to $c_3$ can be implemented rather elegantly, but it is the terms proportional to $c_+$, stemming from a tensor one-loop triangle integral of rank 4, that make the generic result (including the unitarity corrections) lengthy, and additional form factors appear compared to \cref{eq:Kpigg_LO}.
Consequently, going beyond the LO, the complete result employing the ansatz \eqref{eq:KpiPP_NLO_sum} can be written as
\begin{align}
&\mathcal{M}_{\rho\sigma}\big(K(P)\to\pi(r)\gamma_\rho^*(k_1)\gamma_\sigma^*(k_2)\big)\notag\\
&=\xi e^2\Big\{F_{\rho\sigma}\bigl(k_1,k_2\bigr)
+G\big(k_1^2,k_2^2\big)T_{\rho\sigma}^{(2)}(k_1,k_2)\Big\}\notag\\
&\;\;+e^2\GF M_K^2
\Big\{
    \Big[a^{(1)}(z,z_0)+\frac\kappa{z_0}\sum_P\bar A^{(1)}\Big]T_{\rho\sigma}^{(2)}(k_1,k_2)\notag\\
    &\quad\;\;+\Big(a^{(2)}+\frac\kappa{z_0}\sum_P\bar A^{(2)}\Big)T_{\rho\sigma}^{(4)}(k_1,k_2)\notag\\
    &\quad\;\;+\Big(a^{(3)}+\frac\kappa{z_0}\sum_P\tilde c_+A_{c_+}^{(3)}\Big)\big[T_{\rho\sigma}^{(4)}(P,P)+T_{\rho\sigma}^{(4)}(r,r)\big]\notag\\
    &\quad\;\;+\Big[a^{(4)}+\frac\kappa{z_0}\sum_P\tilde c_+A_{c_+}^{(4)}(k_1,k_2)\Big]T_{\rho\sigma}^{(4)}(k_1,P+r)\notag\\
    &\quad\;\;+\Big[a^{(4)}+\frac\kappa{z_0}\sum_P\tilde c_+A_{c_+}^{(4)}(k_2,k_1)\Big]T_{\rho\sigma}^{(4)}(P+r,k_2)
\Big\}\notag\\
&+C_1T_{\rho\sigma}^{(a)}
+C_2T_{\rho\sigma}^{(b)}
\,,
\label{eq:Kpigg_NLO}
\end{align}
The contribution related to topology (1) is represented by the first line on the right-hand side of \cref{eq:Kpigg_NLO}, with $F_{\rho\sigma}\bigl(k_1,k_2\bigr)$ having the same form as in \cref{eq:F_rs,eq:Ftilde_rs} while employing $F(q^2)$ from \cref{eq:F_gen} and with
\begin{equation}
\frac1{\GF M_K^2}G(k_1^2,k_2^2)
    =\frac1\GF\bigl[F(k_1^2)+F(k_2^2)\bigr]_{b_+=\,0}-4d^{(4)}\,.
\end{equation}
This is followed by the contribution associated with topology (2a), which appears summed over all considered pseudoscalars $P$ running in the loop (representing the contributing channels) and accompanied by the unconstrained (subtraction) polynomial part, which is specified later.
Above, I further used $\tilde c_+=c_+/r_\chi^2$, and
\begin{equation}
\begin{aligned}
    \bar A^{(1)}&=j_P\hat A^{(1)}+\tilde c_+A_{c_+}^{(1)}-\frac{z_1z_2}{z_0}\bar A^{(2)}\,,\\
    \frac{\lambda_z}{4z_0}\bar A^{(2)}&=j_P\hat A^{(2)}+\tilde c_+A_{c_+}^{(2)}\,.
\end{aligned}
\label{eq:Kpigg_NLO_A1A2}
\end{equation}
(For compactness, I switched to dimensionless variables $z_{1,2}=k_{1,2}^2/M_K^2$, $z_0=k_1\cdot k_2/M_K^2$, and $\lambda_z=\lambda_s/M_K^4$ used also further, especially in \cref{sec:Ai}.)
The amplitudes $\hat A^{(1)}$ and $\hat A^{(2)}$ are from \cref{eq:A1_LO,eq:A2_LO};
the additional factors related to $\hat A^{(2)}$ and how it enters \cref{eq:Kpigg_NLO} stem from switching the basis structure $\hat T_{\rho\sigma}^{(2)}$ [used in \cref{eq:Kpigg_LO}] to $T_{\rho\sigma}^{(4)}(k_1,k_2)$.
The tensor-structure functions $T_{\rho\sigma}^{(2)}$ and $T_{\rho\sigma}^{(4)}$ are defined in \cref{sec:structures}, $T_{\rho\sigma}^{(a)}$ in \cref{eq:structs}, and
\begin{multline}
	M_K^4T_{\rho\sigma}^{(b)}
	=\varepsilon_{\;\;\sigma\alpha\beta}^{\kappa}k_2^\alpha P^\beta\big(g_{\rho\kappa}P\cdot k_1-P_\rho k_{1\kappa}\big)\\
    +\{k_1\leftrightarrow k_2,\rho\leftrightarrow\sigma\}
\end{multline}
is the remaining piece of the anomalous part [topology (2b)], which is not discussed further here.
This completes the (manifestly gauge-invariant) Lorentz structure of the doubly off-shell amplitude.
The four new subamplitudes $A_{c_+}^{(i)}$ are listed in \cref{sec:Ai}.
At LO in ChPT, $\frac1\GF F(q^2)\big|_{b_+=\,0}=-c^{(4)}$, and $c^{(4)}$ and $d^{(4)}$ are given in \cref{eq:c/d(4)+,eq:c/d(4)0} of \cref{sec:CTs}.
One then arrives at
$\dfrac1{\GF M_K^2}G(k_1^2,k_2^2)=-2c_+^{(4)}-4d_+^{(4)}=-2\go\kappa\hat c$,
a finite combination of LECs; cf.~\cref{eq:Kpigg_LO}.

For completeness and consistency at the given order, one should consider the meson--photon interactions beyond the minimal coupling.
This can be done by multiplying the appropriate parts of the amplitude by EM form factors~\cite{DAmbrosio:1998gur}.

While the UV-finite part introduces some complexities, treating UV-divergent terms remains straightforward.
Regarding topology (1), $F_{\rho\sigma}(k_1,k_2)$ is already finite, as it is built from renormalized one-photon transition form factors $F(q^2)$, and only about the $G$-dependent remainder proportional to $T_{\rho\sigma}^{(2)}(k_1,k_2)$ does one need to be careful.
[This term is connected to the extra piece \eqref{eq:L2_kappa} of \cref{sec:EFT}.]
The 1PI part is then finite at LO, and beyond that, only the term proportional to $c_+$ contains UV-divergent integrals.
The UV-divergent part of the 1PI amplitude (before the renormalization) reads
\begin{equation}
\begin{aligned}
    &\mathcal{M}_{\rho\sigma}^\text{(1PI, UV)}\big(K(P)\to\pi(r)\gamma_\rho^*(k_1)\gamma_\sigma^*(k_2)\big)\\
    &=e^2\GF M_K^2\sum_Pc_+^{(P)}\frac23\frac{M_K^2}{\ChPT^2}
    \bigg(\frac\kappa2\frac1{\tilde\epsilon}-\frac12L_P\bigg)\\
    &\times\Bigl\{
    \frac1{M_K^2}\big(P\cdot r+k_1\cdot k_2-P^2-r^2\big)T_{\rho\sigma}^{(2)}(k_1,k_2)\\
    &\quad-T_{\rho\sigma}^{(4)}(k_1,k_2)
    +\big[T_{\rho\sigma}^{(4)}(P,P)+T_{\rho\sigma}^{(4)}(r,r)\big]
    \Big\}\,,
\end{aligned}
\label{eq:Mrs_1PI_UV}
\end{equation}
which, unsurprisingly, can be matched by the polynomial structure of the CT amplitude~\eqref{eq:M_CT(6)}.
Thus, the latter can effectively be replaced by the associated polynomial of its finite parts, accompanied by the logarithms $L_P$, leading to a scale-independent result.
In particular, the unconstrained polynomials of \cref{eq:Kpigg_NLO} are
\begin{equation}
\begin{aligned}
    &{r_\chi^2}a^{(1)}(z,z_0)
    =-\cBr_1+L(1+r_\pi^2)-\tfrac12\cBr_5(1-r_\pi^2)\\[-1mm]
    &\;\;-\tfrac12\big(L+\cBr_2\big)(1+r_\pi^2-z)-\big(L-\cBr_3\big)z_0\,,\\[1mm]
    &{r_\chi^2}a^{(2)}=+L+\cBr_6\,,\\
    &{r_\chi^2}a^{(3)}=-L-\cBr_4\,,\\
    &{r_\chi^2}a^{(4)}=-\tfrac12\cBr_5\,,
\end{aligned}
\label{eq:Kpigg_pol}
\end{equation}
with $L=\sum_P\tfrac13c_+^{(P)}L_P$, which is an analog to \cref{eq:F_pol}.

Finally, although the additional subamplitudes $A_{c_+}^{(i)}$ for the doubly off-shell case are somewhat tedious, the on-shell case can be presented in a rather compact form, reducing the expressions from \cref{sec:Ai} to
\begin{gather}
\begin{aligned}
    A_{c_+}^{(1)}&(z,y)
    =\frac16\big(y_1^2+y_2^2\big)\Big(1-\frac3{\bar z_0}\Jbar_P\Big)\\
    &+r_P^2z_0\Bigl\{-\tfrac89\bar z_0
    +y_t\big(1-\tfrac5{18}\bar z_0\big)
    +\tfrac12y_p(2+\bar z_0)\\
    &+\tfrac13\Jbar_P\big[2-\bar z_0+6y_p+y_t(1+\bar z_0)\big]\\
    &+2M_P^2C_0\big[y_t-\bar z_0+y_p(1+\bar z_0)\big]\Big\}\,,
\end{aligned}\\[2mm]
\begin{aligned}
    A_{c_+}^{(3)}(z)
    =r_P^2\big[\tfrac1{18}\bar z_0+\tfrac13(\bar z_0-5)\Jbar_P-\big(1+2M_P^2C_0\big)\big]\,,
\end{aligned}
\end{gather}
where $\Jbar_P\equiv\Jbar_P(zM_K^2)$, $C_0=C_0(zM_K^2,0,0;M_P^2)$, $z=2z_0$, $z_0=k_1\cdot k_2/M_K^2$, $\bar z_0=z_0/r_P^2$, $y_t=\frac12(1-y_0/z)$, $y_p=4y_1y_2/z^2$, with $y_0=2(1+r_\pi^2-z_0)$ and $y_{1,2}=\frac12(1-r_\pi^2\pm y)$.
Subamplitudes $A_{c_+}^{(2)}$ and $A_{c_+}^{(4)}$ do not contribute.
For the definitions of the loop functions, see \cref{sec:functions}.

\section{Summary and outlook}
\label{summary}

In this work, the calculation of the $K\to\pi\gamma^*\gamma^*$ transition amplitudes was presented in several stages.
First, the LO ChPT result was shown in \cref{sec:2g}, and \cref{eq:Kpigg_LO} supplemented by the subamplitudes \eqref{eq:Ai+}--\eqref{eq:A2_LO} represents a novel result itself.
To improve its phenomenological relevance, this was further extended with unitarity corrections based on a phenomenological $K\to3\pi$ amplitude, as it is reasonable to assume the two-pion rescattering is a dominant effect beyond LO.
Including the quadratic fit of the $K\to3\pi$ Dalitz plot also affects the $\mathcal{O}(p^2)$ part of the amplitude, so a calculation with generic coefficients going beyond the LO ChPT result is necessary.
This is presented in \cref{sec:general,sec:gen_res_NLO}.

Out of the three new $\mathcal{O}(p^4)$ parameters introduced in \cref{eq:K3pi_NLO_gen}, the terms proportional to $c_3$ and $c_-$ have a rather limited impact on the overall form of the amplitude, but $c_+$ leads to new (remaining) gauge-invariant structures and renders the result \eqref{eq:Kpigg_NLO} significantly less compact; see \cref{sec:Ai} for the additional subamplitudes.
Besides that, a nonvanishing $c_+$ also spoils some straightforward properties of the tadpole contribution \eqref{eq:tadpole_gen} discussed in \cref{sec:tadpole_gen}.
Regarding the different coefficients and terms of the ansatz \eqref{eq:KpiPP_NLO_sum}, only those proportional to $b_+$ and $c_-$ enter the one-photon transition, as it is seen in \cref{eq:h(q^2),eq:hhat(q^2),eq:F_gen,eq:F_pol}.
On the other hand, the only coefficient on which the 1PI part of the two-photon transition does not depend is $c_-$.

The general ansatz \eqref{eq:KpiPP_NLO_sum} also allows one to understand the relation among the charged and neutral channels, which are treated in parallel and in a rather general way throughout the work.
The LO ChPT calculation already exhibits the factorizations \eqref{eq:Ai+} and \eqref{eq:Ai0}, holding for both form factors in the doubly off-shell case.
The subsequent generalized treatment leads to the expression in \cref{eq:jP} that allows for understanding these factors, and the LO ChPT results can be obtained by substituting the coefficients from \cref{tab:amplitudes} in \cref{sec:amplitudes} into \cref{eq:jP}.

A direct LO ChPT calculation based on a standard set of diagrams ({\em not} employing the notation of \cref{sec:notation}), performed in parallel using two parametrizations \eqref{eq:params} provides only limited insight into the underlying structure of the amplitude.
Although the square-root parametrization simplifies the calculation by rendering some of the radiative vertices vanishing, all the remaining tadpole diagrams, for instance, must be considered to obtain a gauge-invariant result.
This demonstrates a common feature of the standard approach:
Gauge invariance is only recovered after all the diagrams are summed.
In contrast, the methods utilized in this work provide a more intuitive alternative.
Separating the diagrams (and consequently the amplitude) into naturally gauge-invariant parts using the notation described in detail in \cref{sec:notation} allows for simplifications, as some of the subsets are vanishing.
These features propagate to the generic case.
Needless to say, plugging \cref{fig:K->pig_LO_ChPT_a,fig:K->pigg_LO_ChPT} into \cref{fig:K->pigg} leads to an equivalent set of diagrams as plugging \cref{fig:K->pig_LO_ChPT_b,fig:K->pigg_LO_1PI,fig:K->pigg_LO_1PR} into \cref{fig:K->pigg_LO}, once \cref{fig:KpiPPg(g)} is considered.

Finally, allowing for additional free parameters related to different off-shell extensions lets one identify the physically relevant contributions.
At LO, the tadpole related to the $K\to\pi$ weak mixing vanishes for a specific choice of these parameters.
Consequently, all the tadpole diagrams in $K^+\to\pi^+\gamma^*(\gamma^*)$ vanish.
Hence, in such a parametrization, many diagrams can be ignored.
Furthermore, choosing the described ``partial'' way of shifting the vertex momenta by photon polarization vectors and its diagrammatic representation leads, on-shell, to separating diagrams into gauge-invariant subsets.
And tadpole contributions form some of them.
Thus, the concepts of associating contributions related by gauge invariance and studying the off-shell extrapolations in the given theoretical framework at given orders lead to significant simplifications and to a more efficient and physically transparent calculation.
Regarding the general structure, the difference between the charged and neutral channels is then given by the 1PR diagrams stemming from the 1PI part of the $K\to\pi\gamma^*$ amplitude.

The generic case not only allows the recovery of the LO ChPT result \eqref{eq:Kpigg_LO} based on the octet Lagrangian but, in addition, the 27-plet terms can be obtained by using the corresponding LO ChPT $K\pi PP$ amplitudes, i.e., including the $G_{27}$ terms.
Finally, the results can be utilized to obtain contributions to decays of pseudoscalars other than kaons, e.g., $\eta^{(\prime)}$.
This can be achieved by setting the masses and parameters of the central $P_1P_2P\bar P$ vertex appropriately, replacing the $K\pi P\bar P$ amplitude, as the initial-state kaon mass $M_K$ and the final-state pion mass $M_\pi$ are distinguished from the loop mass $M_P$.

It is important to emphasize again that the presented results do not represent a full-scale NLO (two-loop) computation.
The way the result \eqref{eq:Kpigg_NLO} should be used is as follows.
First, one identifies the contributing channels to a given process, e.g., $K\to\pi+(\pi\pi/KK)\to\pi+\gamma^*\gamma^*$.
Then the parametrizations \eqref{eq:KpiPP_NLO_sum} of the contributing channel vertices, e.g., $K\pi\pi\pi$ and $K\pi KK$, need to be fixed.
The parameters then enter \cref{eq:jP}, and the resulting $A^{(i)}$ and $c_+$ enter \cref{eq:Kpigg_NLO}.
If the decaying particle is charged, then the contribution of the one-photon transition \eqref{eq:F_gen}, e.g., $K^+\to\pi^+\gamma^*$, is added in an analogous way in terms of $F_{\rho\sigma}$ and $G$.
Alternatively, a phenomenological $F(q^2)$ can be used in $F_{\rho\sigma}$.
The amplitude is finalized by adding an unconstrained polynomial stemming from the allowed tree-level CT contributions and related renormalization procedure.
Its coefficients need to be established from fits to data, or, e.g., by studying the resonance contributions.
Alternatively, the LECs' finite parts can be put to zero and vary within natural bounds establishing model-independent determinations.

This work represents a baseline for further investigation into the processes already mentioned in the Introduction.
In the works that will follow, I first plan to write up a phenomenological study of the $K^+\to\pi^+\gamma\ell^+\ell^-$ decays that improves on Ref.~\cite{Gabbiani:1998tj}, among others, by discussing the relevance of the (to-a-large-extent) ignored 1PR contributions (in the sense of \cref{fig:K->pigg}).
This is motivated by already ongoing analyses of these modes at NA62.
Second, the study of the consequences of the present work on $K\to\pi\ell_1^+\ell_1^-\ell_2^+\ell_2^-$ decays will be performed: In Ref.~\cite{Husek:2022vul}, the unknown 1PI doubly off-shell form factors were ignored beyond a constant term equivalent to $\hat c$; see also \cref{sec:EFT}.
These channels are interesting as their improved branching-ratio determinations are relevant for searches beyond the Standard Model.
Finally, the effects on radiative corrections in $K\to\pi\ell^+\ell^-$ decays will be considered.
These are essential for ongoing $K^+\to\pi^+\ell^+\ell^-$ form-factor measurements at NA62 in both the electron and muon channels.
A significant statistical improvement compared to the latest NA62 result from $K^+\to\pi^+\mu^+\mu^-$~\cite{NA62:2022qes} is foreseen, as the entire final dataset will be analyzed.
This, in turn, needs to be complemented by correspondingly precise inputs for the Monte Carlo decay generator.

\begin{acknowledgments}

I thank Johan Bijnens, Evgueni Goudzovski, Karol Kampf, Stefan Leupold, and Ji\v r\'i Novotn\'y for discussions in various stages of the calculation.

This work was supported by the MSCA Fellowships CZ--UK2 project No.~CZ.02.01.01/00/22\_010/0008115 financed by Programme Johannes Amos Comenius (OP JAK).

\end{acknowledgments}


\appendix

\section{Radiative vertices and diagrammatic notation}
\label{sec:notation}

\begin{figure}[t]
\centering
\includegraphics[width=0.98\columnwidth]{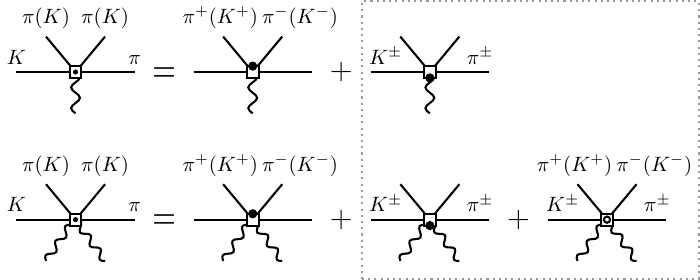}
\caption{
    Radiative vertices $K\pi PP\gamma$ and $K\pi PP\gamma\gamma$ and their separation into ``internal (I)'' and ``external (E)'' parts, stemming from the minimal-coupling extension of the $K\pi PP$ vertex.
    Using this diagrammatic notation, subsets of gauge-invariant diagrams can be identified.
    On the left-hand sides, the combination of weak and (minimal) electromagnetic couplings is naturally (see notation in other figures) represented by a square with a centered dot.
    Depending on which meson-field pair the covariant derivatives act on (internal or external), this dot is shifted (up or down, respectively).
    As denoted by the leg labels, the respective sets of mesons need to be charged; otherwise, such a subdiagram vanishes.
    In the case of the two-photon transition and when all the mesons are charged, there is a remainder, denoted by an empty circle centered inside the square.
}
\label{fig:KpiPPg(g)}
\end{figure}

In this appendix, the matrix elements for the radiative vertices used in this work [based on the generic matrix element \eqref{eq:KpiPP_NLO_sum}] are explicitly listed and related to the diagrammatic representation of \cref{fig:KpiPPg(g)}; see also its detailed caption.
This notation is designed to group diagrams into gauge-invariant subsets, each corresponding to a distinct flow of electric charge.
Specifically, diagrams with external- and internal-leg radiation must be grouped with diagrams containing their corresponding radiative contact-term vertices.
The inclusion of these contact terms is required to satisfy the Ward identity for each of these currents.
These terms are generated from the external (internal) momenta of the charged particles flowing into the vertex, which can be calculated either by shifting these momenta or, equivalently, by taking the derivatives with respect to them, as presented later in \cref{eq:M(EI)_r_D,eq:M(EI)_rs_D}.

Employing the shorthand notation for the combinations of (to-be) internal (loop) $\qIPM=p_1\pm p_2$ and external $\qEPM=P\pm r$ meson 4-momenta, the amplitudes obtained by {\em gradually} replacing the partial derivatives $\partial_\mu$ by covariant derivatives $D_\mu$ acting on (to-be) external (internal) fields in the corresponding Lagrangian [leading to \cref{eq:KpiPP_NLO_sum}] are as follows.
For the case with one additional photon,
\begin{widetext}
\begin{subequations}
\label{eq:M_KpiPPg}
\begin{align}
    \frac1{e\GF}\mathcal{M}_{\text{(E)},\,\rho}^\text{(2+4)}\bigl(K^+(P)\pi^-(r)P(p_1)\bar P(p_2)\gamma_\rho(k)\bigr)
    &
    =2a_K P_\rho-2a_\pi r_\rho+(a_K-a_\pi)k_\rho
    +b_+\qIM_\rho
    \notag\\
    &\hspace{-2cm}
    -c_3\qEM_\rho\frac{p_1\cdot p_2}{\ChPT^2}
    -c_+\qEM_\sigma\frac{(p_{1\rho}p_{2}^{\sigma}+p_{1}^{\sigma}p_{2\rho})}{\ChPT^2}
    +c_-\qEP_\sigma\frac{(p_{1\rho}p_{2}^{\sigma}-p_{1}^{\sigma}p_{2\rho})}{\ChPT^2}\,,
    \label{eq:M_KpiPPg_E}\\
    \frac1{e\GF}\mathcal{M}_{\text{(I)},\,\rho}^\text{(2+4)}\bigl(K(P)\pi(r)P^+(p_1)P^-(p_2)\gamma_\rho(k)\bigr)
    &
    =2a_1 p_{1\rho}-2a_2 p_{2\rho}+(a_1-a_2)k_\rho
    +b_+\qEM_\rho
    -b_0\qIM_\rho
    \notag\\
    &\hspace{-2cm}
    -c_3\qIM_\rho\frac{P\cdot r}{\ChPT^2}
    -c_+\qIM_\sigma\frac{(P_{\rho}r^{\sigma}+P^{\sigma}r_{\rho})}{\ChPT^2}
    +c_-\qIP_\sigma\frac{(P_{\rho}r^{\sigma}-P^{\sigma}r_{\rho})}{\ChPT^2}\,,
    \label{eq:M_KpiPPg_I}
\end{align}
\end{subequations}
and for the case with two additional photons,
\begin{subequations}
\label{eq:M_KpiPPgg}
\begin{align}
    \frac1{e^2\GF}\mathcal{M}_{\text{(E)},\,\rho\sigma}^\text{(2+4)}\bigl(K^+(P)\pi^-(r)P(p_1)\bar P(p_2)\gamma_\rho^*\gamma^*_\sigma\bigr)
    &=2g_{\rho\sigma}\bigg(a_K+a_\pi-c_3\frac{p_1\cdot p_2}{\ChPT^2}\bigg)
    -2c_+\frac{(p_{1\rho}p_{2\sigma}+p_{1\sigma}p_{2\rho})}{\ChPT^2}\,,
    \label{eq:M_KpiPPgg_E}\\
    \frac1{e^2\GF}\mathcal{M}_{\text{(I)},\,\rho\sigma}^\text{(2+4)}\bigl(K(P)\pi(r)P^+(p_1)P^-(p_2)\gamma_\rho^*\gamma^*_\sigma\bigr)
    &=2g_{\rho\sigma}\bigg(a_1+a_2-b_0-c_3\frac{P\cdot r}{\ChPT^2}\bigg)
    -2c_+\frac{(P_\rho r_\sigma+P_\sigma r_\rho)}{\ChPT^2}\,,
    \label{eq:M_KpiPPgg_I}\\
    \frac1{e^2\GF}\mathcal{M}_{\text{(R)},\,\rho\sigma}^\text{(2+4)}\bigl(K^+(P)\pi^-(r)P^+(p_1)P^-(p_2)\gamma_\rho^*\gamma^*_\sigma\bigr)
    &=2g_{\rho\sigma}\bigg(2b_++c_+\frac{\qEM\cdot \qIM}{\ChPT^2}
    +c_-\frac{\qEP\cdot \qIP}{\ChPT^2}\bigg)\notag\\
    &\hspace{-2cm}+(c_3+c_+)\frac{\qEM_\rho \qIM_\sigma+\qEM_\sigma \qIM_\rho}{\ChPT^2}
    -c_-\frac{\qEP_\rho \qIP_\sigma+\qEP_\sigma \qIP_\rho}{\ChPT^2}\,.
    \label{eq:M_KpiPPgg_R}
\end{align}
\end{subequations}
\end{widetext}
The convention is such that all momenta are incoming.
Diagrammatically, this is represented in \cref{fig:KpiPPg(g)}.
The first diagrams on the right-hand side correspond to the parts of the amplitude labeled as ``(I),'' the second diagrams represent the ``(E)'' part, and the remainder ``(R)'' is represented by the very last subdiagram.

Inspecting the generic matrix element \cref{eq:KpiPP_NLO_sum}, all the terms except those proportional to the four coefficients $a_i$ are at most linear in all 4-momenta.
This means that these terms are associated with Lagrangian monomials with at most one partial derivative acting on each field.
Consequently, the above matrix elements can be obtained in terms of derivatives with respect to the (to-be) external (E) or internal (I) momenta, taking into account meson charges $\eta_P$.
(This represents an alternative to the procedure consisting of a shift of momenta by polarization vectors corresponding to the replacement of partial derivatives by covariant ones.)
In particular, using
\begin{subequations}
\begin{align}
    \partial_\rho^\text{(E)}&\equiv \eta_P\frac{\partial}{\partial P^\rho}+\eta_r\frac{\partial}{\partial r^\rho}\,,\\
    \partial_\rho^\text{(I)}&\equiv \eta_{p_1}\frac{\partial}{\partial p_1^\rho}+\eta_{p_2}\frac{\partial}{\partial p_2^\rho}\,,
\end{align}
\end{subequations}
one can write (including corrections due to the mentioned quadratic terms)
\begin{subequations}
\label{eq:M(EI)_r_D}
\begin{align}
    \mathcal{M}_{\text{(E)},\,\rho}^\text{(2+4)}
    &=e\bigg[\partial_\rho^\text{(E)}+k_\rho\bigg(\eta_P\frac{\partial}{\partial P^2}+\eta_r\frac{\partial}{\partial r^2}\bigg)\bigg]\mathcal{M}_{K\pi PP}^\text{(2+4)}\,,\\
    \mathcal{M}_{\text{(I)},\,\rho}^\text{(2+4)}
    &=e\bigg[\partial_\rho^\text{(I)}+k_\rho\bigg(\eta_{p_1}\frac{\partial}{\partial p_1^2}+\eta_{p_2}\frac{\partial}{\partial p_2^2}\bigg)\bigg]\mathcal{M}_{K\pi PP}^\text{(2+4)}\,,
\end{align}
\end{subequations}
and finally,
\begin{subequations}
\label{eq:M(EI)_rs_D}
\begin{align}
    \mathcal{M}_{\text{(E)},\,\rho\sigma}^\text{(2+4)}
    &=e\partial_\sigma^\text{(E)}\mathcal{M}_{\text{(E)},\,\rho}^\text{(2+4)}=e^2\partial_\rho^\text{(E)}\partial_\sigma^\text{(E)}\mathcal{M}_{K\pi PP}^\text{(2+4)}\,,\\
    \mathcal{M}_{\text{(I)},\,\rho\sigma}^\text{(2+4)}
    &=e\partial_\sigma^\text{(I)}\mathcal{M}_{\text{(I)},\,\rho}^\text{(2+4)}
    =e^2\partial_\rho^\text{(I)}\partial_\sigma^\text{(I)}\mathcal{M}_{K\pi PP}^\text{(2+4)}\,,\\
    \mathcal{M}_{\text{(R)},\,\rho\sigma}^\text{(2+4)}
    &
    =e\partial_\sigma^\text{(I)}\mathcal{M}_{\text{(E)},\,\rho}^\text{(2+4)}
    +e\partial_\sigma^\text{(E)}\mathcal{M}_{\text{(I)},\,\rho}^\text{(2+4)}\notag\\
    &=e^2\big[\partial_\sigma^\text{(I)}\partial_\rho^\text{(E)}
    +\partial_\sigma^\text{(E)}\partial_\rho^\text{(I)}\big]
    \mathcal{M}_{K\pi PP}^\text{(2+4)}\,,
\end{align}
\end{subequations}
in which the terms proportional to $k=-P-r-p_1-p_2$ dropped out due to opposite charges of the meson pairs.

To compare the results \eqref{eq:M_KpiPPg} and \eqref{eq:M_KpiPPgg} with literature, one expects the need to sum the respective subamplitudes in \cref{eq:M_KpiPPg} and (separately) those in \cref{eq:M_KpiPPgg}, as the respective contributions are typically treated as one single vertex.
This applies except in the case of the neutral channel, in which only the internal parts survive.
For instance, some NLO terms of \cref{eq:M_KpiPPg_I} should match Eq.~(20) of Ref.~\cite{Donoghue:1997rr}, and some NLO terms in \cref{eq:M_KpiPPgg_I} should match Eq.~(21) of the cited work, taking $c_3\GF/\ChPT^2=4a_1$ and $c_+\GF/\ChPT^2=4a_2$; this follows from comparing \cref{eq:K3pi_NLO_gen} with Eq.~(18) of Ref.~\cite{Donoghue:1997rr}.
Unfortunately, there seem to be some obvious typos in Ref.~\cite{Donoghue:1997rr} regarding dimensions (a missing scalar product) and signs.

\section{Definitions of used (loop) functions}
\label{sec:functions}

The functions representing the results are closely related to the standard Passarino--Veltman one-loop integrals $A_0$, $B_0$, and $C_0$~\cite{tHooft:1978jhc,Passarino:1978jh}.
In what follows, I use the compact notation for the Feynman denominators with loop momentum $\ell$ and mass $M$,
\begin{equation}
D(q) \equiv(\ell-q)^2-M^2+i\epsilon\,.
\end{equation}
The scalar one-loop integrals read
\begin{align}
\kappa A_0(M^2)
&=\frac1i\!\int\!\frac{\text{d}^d\ell}{(2\pi)^d}\frac{1}{D(0)}\,,\\
\kappa B_0(q^2;M^2)
&=\frac1i\!\int\!\frac{\text{d}^d\ell}{(2\pi)^d}\frac{1}{D(0)D(q)}\,,\label{eq:B0}\\
\kappa C_0(p^2,q^2,(p+q)^2;M^2)
&=\frac1i\!\int\!\frac{\text{d}^d\ell}{(2\pi)^d}\frac{1}{D(0)D(p)D(-q)}\,,
\end{align}
where I used a shorthand notation $B_0(s;M^2)\equiv B_0(s,M^2,M^2)$ and $C_0(s_1,s_2,s_3;M^2)\equiv C_0(s_1,s_2,s_3,M^2,M^2,M^2)$.
The integrals with vanishing momenta are
\begin{align}
B_0(0;M^2)
&=\frac1{i\kappa}\!\int\!\frac{\text{d}^d\ell}{(2\pi)^d}\frac{1}{D^2(0)}
=\frac1{\tilde\epsilon}-1-\log\frac{M^2}{\mu^2}\,,\\
A_0(M^2)
&=M^2\big[1+B_0(0;M^2)\big]\,.
\end{align}
Others with nontrivial momentum dependence are
\begin{align}
B_0(q^2;M_P^2\bigr)
&=B_0(0;M_P^2\bigr)+\Jbar_P(q^2)\,,\label{eq:B0_Jbar}\\
C_0\big(s,0,0;M_P^2\big)
&=\frac{1}{2(s-4M_P^2)}\bigl[\Jbar_P(s)-2\big]^2\,,\\
C_0\big(s_1,s_2,0;M_P^2\big)\notag\\
&\hspace{-2cm}=\frac{s_1C_0\big(s_1,0,0;M_P^2\big)-s_2C_0\big(s_2,0,0;M_P^2\big)}{s_1-s_2}\,.
\end{align}
Above,
\begin{equation}
\Jbar_P(s)
=2+\beta_P(s)\log\biggl(-\frac{1-\beta_P(s)}{1+\beta_P(s)}+i\epsilon\biggr)\,,
\label{eq:Jbar}
\end{equation}
with $\beta_P(s)=\sqrt{1-{4M_P^2}/s}$, and I also use
\begin{equation}
\Jtilde_P\big(s_1,s_2\big)
\equiv\Jbar_P(s_1)-\Jbar_P(s_2)\,.
\end{equation}
The most involved triangle function used, $C_0\big(s_1,s_2,s_3;M^2\big)$, can be expressed in terms of 12 dilogarithms and can be found, e.g., in Eq.~(4.26) of Ref.~\cite{Denner:1991kt}.
In the case relevant to the present work (real parameters and $\lambda>0$), this can be significantly reduced to
\begin{multline}
C_0\big(s_1,s_2,s_3;M_P^2\big)\\
=\!\!\!\!\sum_{\substack{\{i,j,k\}=\{1,2,3\}\\ \quad\qquad(\text{cycl.})}}\;
\sum_{\substack{\sigma=\pm1\\ \sigma'=\pm1}}
\frac\sigma{\sqrt\lambda}\,\text{Li}_2\frac{y_{ijk}+\sigma}{y_{ijk}+\sigma'\widetilde\beta_P(s_k)}\,,
\end{multline}
with $\lambda=\lambda(s_1,s_2,s_3)$,
$y_{ijk}=\dfrac1{\sqrt\lambda}\big(s_i+s_j-s_k\big)$,
and $\widetilde\beta_P(s)=\big[\beta_P(s)+i\epsilon s\big](1+i\epsilon s)$.
This slightly modified definition of $\widetilde\beta_P$ also accounts for cases with $s_i=4M_P^2$ [and thus $\beta_P(s_i)=0$] for some $i=1,2,3$.

The relation of the functions listed above to those used, for instance, in Refs.~\cite{Ecker:1987fm,Cohen:1993ta} is given by
\begin{equation}
\begin{aligned}
    R(z)&=-\tfrac16+\tfrac1z \Jbar_P(M_P^2z)\,,\\
    F(z)&=1+2M_P^2C_0(M_P^2z,0,0;M_P^2)\,;
\end{aligned}
\end{equation}
cf.~\cref{eq:Agg(1)}.

\section{LO ChPT Lagrangian for \texorpdfstring{$K\pi PP$}{KpiPP} vertices}
\label{sec:Lagrangians}

To calculate the considered processes in ChPT, the Lagrangian~\eqref{eq:L_ChPT} needs to be expanded for obtaining the $K\pi PP$ vertices, in particular $K^+\pi^-\pi^+\pi^-$, $K^+\pi^-K^+K^-$, $K^0\pi^0\pi^+\pi^-$, and $K^0\pi^0K^+K^-$.
The vertices $K\pi\pi^0\pi^0$ and $K\pi K^0\overline{K^0}$ are not relevant at considered orders.
Thus, for simplicity, they are not listed below in the form of interaction Lagrangians, but only, for completeness, as on-shell amplitudes in \cref{sec:amplitudes}.
In the exponential [\cref{eq:param_exp}] and square-root [\cref{eq:param_sqrt}] parametrizations, I obtained, respectively,
\begin{widetext}
\begin{align}
\begin{split}
    &\Lb_{\mathcal{O}(p^2)}^\text{(exp)}
    =\Bigl\{-G_8D_\mu K^+D^\mu\pi^+(\pi^-)^2
    +\tfrac13G_8K^+\pi^-\big[\pi^+D^2\pi^--2\pi^-D^2\pi^+-M_\pi^2\pi^+\pi^-\big]\Bigr\}
    +\{K\leftrightarrow\pi\}+\text{h.c.}\\
    &\quad+\tfrac1{\sqrt{2}}G_8K^0\pi^0\bigl[-2D_\mu\pi^+D^\mu\pi^-+M_\pi^2\pi^+\pi^-\big]
    -\tfrac1{3\sqrt{2}}G_8K^0\pi^0\big[\pi^+D^2\pi^-+\pi^-D^2\pi^++2M_\pi^2\pi^+\pi^-\big]
    +\text{h.c.}\\
    &\quad+{\sqrt{2}}G_8K^+\pi^0\bigl[-2\partial_\mu K^0D^\mu K^-+M_K^2K^0K^-\big]\\
    &\quad\qquad-\tfrac{\sqrt{2}}{3}G_8K^+\pi^0\big[K^-\partial^2K^0+K^0D^2K^-+2M_K^2K^0K^-\big]
    +\tfrac{\sqrt{2}}{3}G_8K^0\pi^0\big[K^-D^2K^+-K^+D^2K^-\big]
    +\text{h.c.},
\end{split}
\label{eq:L_O(p2)_exp}
\end{align}
\begin{align}
\begin{split}
    &\Lb_{\mathcal{O}(p^2)}^\text{(sqrt)}
    =\Bigl\{-G_8D_\mu K^+D^\mu\pi^+(\pi^-)^2
    +G_8K^+\pi^-\big[\pi^+D^2\pi^-+M_\pi^2\pi^+\pi^-\big]\Bigr\}
    +\{K\leftrightarrow\pi\}+\text{h.c.}\\
    &\quad+\tfrac1{\sqrt{2}}G_8K^0\pi^0\bigl[-2D_\mu\pi^+D^\mu\pi^-+M_\pi^2\pi^+\pi^-\big]
    -\tfrac1{\sqrt{2}}G_8K^0\pi^0\big[\pi^+D^2\pi^-+\pi^-D^2\pi^++2M_\pi^2\pi^+\pi^-\big]
    +\text{h.c.}\\
    &\quad+{\sqrt{2}}G_8K^+\pi^0\bigl[-2\partial_\mu K^0D^\mu K^-+M_K^2K^0K^-\big]
    -\tfrac{\sqrt{2}}{3}G_8K^+\pi^0\big[K^-\partial^2K^0+3K^0D^2K^-+4M_K^2K^0K^-\big]
    +\text{h.c.}
\end{split}
\label{eq:L_O(p2)_sqrt}
\end{align}
\end{widetext}
The above expressions should be equivalent to Lagrangians shown, e.g., in Ref.~\cite{Ecker:1987fm}.
Alternatively, for the $K^0\pi^0\pi^+\pi^-$ case in \cref{eq:L_O(p2)_exp} (its second line), one can write
\begin{multline}
    \widetilde{\Lb}_{\mathcal{O}(p^2)}^\text{(exp)}
    \supset\tfrac1{3\sqrt{2}}G_8\bigl[\partial_\mu(K^0\pi^0)\partial^\mu(\pi^+\pi^-)\\
    -K^0\pi^0(4D_\mu\pi^+D^\mu\pi^--M_\pi^2\pi^+\pi^-)\big]
    +\text{h.c.},
\end{multline}
and the last two lines of \cref{eq:L_O(p2)_sqrt} can be written as
\begin{equation}
    \begin{aligned}
    &\widetilde{\Lb}_{\mathcal{O}(p^2)}^\text{(sqrt)}
    \supset\tfrac1{\sqrt{2}}G_8\bigl[\partial_\mu(K^0\pi^0)\partial^\mu(\pi^+\pi^-)\\
    &\qquad\qquad\qquad\quad\;-M_\pi^2K^0\pi^0\pi^+\pi^-\big]
    +\text{h.c.}\\
    &+\tfrac{\sqrt{2}}3G_8\bigl[3D_\mu(K^0K^-)D^\mu(K^+\pi^0)\\
    &\qquad\qquad\;+(2\partial^2K^0-M_K^2K^0)\pi^0K^+K^-\big]
    +\text{h.c.}
    \end{aligned}
\end{equation}

In \cref{eq:L_O(p2)_exp,eq:L_O(p2)_sqrt}, one can identify a handful of the dynamically relevant terms for the considered processes at given order (leading to the correct on-shell amplitudes):
\begin{align}
    &\Lb_{\mathcal{O}(p^2)}
    \supset-G_8D_\mu K^+D^\mu\pi^+(\pi^-)^2\notag\\
    &-G_8D_\mu\pi^-D^\mu K^-(K^+)^2\notag\\
    &+\tfrac1{\sqrt{2}}G_8K^0\pi^0\bigl[-2D_\mu\pi^+D^\mu\pi^-+M_\pi^2\pi^+\pi^-\big]\\
    &+{\sqrt{2}}G_8K^+\pi^0\bigl[-2\partial_\mu K^0D^\mu K^-+M_K^2K^0K^-\big]
    +\text{h.c.}\notag
\end{align}

\section{LO ChPT amplitudes for \texorpdfstring{$K\pi PP$}{KpiPP} vertices}
\label{sec:amplitudes}

In the following results for the $K\pi PP$ vertex amplitudes, I use the convention in which all momenta are incoming and consider the particles on-shell (with $p_{1,2}^2=M_P^2$):
The off-shell extrapolations for all particles are discussed separately in terms of a general off-shell amplitude extension, and it is shown that they do not contribute in the considered cases.

Regarding the charged channel ($K^+\pi^-P\bar P$), I find (couplings $G_8$ are suppressed for compactness, denoting simply $\widetilde{\mathcal{M}}\equiv\frac1{G_8}\mathcal{M}$)
\begin{align}
    \widetilde{\mathcal{M}}^{(2)}\big(K_{(P)}^+\pi_{(r)}^-\pi_{(p_1)}^+\pi_{(p_2)}^-\big)
        &=2P\cdot p_1\,,\label{eq:Kch3pi}\\
    \widetilde{\mathcal{M}}^{(2)}\big(K_{(P)}^+\pi_{(r)}^-K_{(p_1)}^+K_{(p_2)}^-\big)
        &=2r\cdot p_2\,,\label{eq:3Kchpi}\\
    \widetilde{\mathcal{M}}^{(2)}\big(K_{(P)}^+\pi_{(r)}^-\pi_{(p_1)}^0\pi_{(p_2)}^0\big)
        &=-\big[2p_1\cdot p_2+M_\pi^2\big]\,,\\
    \widetilde{\mathcal{M}}^{(2)}\big(K_{(P)}^+\pi_{(r)}^-K_{(p_1)}^0\overline{K^0}_{\!\!\!\!(p_2)}\big)
        &=(r-p_2)\cdot(P-p_1)\,.
\end{align}
For the neutral channel ($K^0\pi^0P\bar P$), I find
\begin{align}
    \widetilde{\mathcal{M}}^{(2)}\big(K^0_{(P)}\pi^0_{(r)}\pi^+_{(p_1)}\pi^-_{(p_2)}\big)
        &=\tfrac1{\sqrt{2}}\big[2p_1\cdot p_2+M_\pi^2\big]\,,\\
    \widetilde{\mathcal{M}}^{(2)}\big(K^0_{(P)}\pi^0_{(r)}K^+_{(p_1)}K^-_{(p_2)}\big)
        &={\sqrt{2}}\big[2P\cdot p_2+M_K^2\big]\,,\\
    \widetilde{\mathcal{M}}^{(2)}\big(K^0_{(P)}\pi^0_{(r)}\pi^0_{(p_1)}\pi^0_{(p_2)}\big)
        &=\tfrac1{\sqrt{2}}M_K^2\,,\\
    \widetilde{\mathcal{M}}^{(2)}\big(K^0_{(P)}\pi^0_{(r)}K^0_{(p_1)}\overline{K^0}_{\!\!\!\!(p_2)}\big)
        &=0\,.
\end{align}
If one defines $t_0\equiv p_1\cdot p_2$ and $t_\pm\equiv(P\cdot p_1\pm r\cdot p_2)$, then, for instance, $2P\cdot p_1=t_++t_-$ and $2r\cdot p_2=t_+-t_-$.
On-shell, $t_-=-\frac12(M_K^2-M_\pi^2)$, so the amplitudes~\eqref{eq:Kch3pi} and~\eqref{eq:3Kchpi} differ just by a constant term.
Also the remaining scalar products (and whole on-shell amplitudes) at $\mathcal{O}(p^2)$ can be brought to the form $aM_K^2+b_0t_0+b_+ t_+$ [cf.\ \cref{eq:KpiPP_LO_gen}],
and the results are listed in \cref{tab:amplitudes}.

\begin{table}[t]
    \centering
    \setlength{\tabcolsep}{9.5pt}
    \renewcommand{\arraystretch}{1.4}
    \begin{tabular}{c | c | c c c}
    \toprule
        Channel & Loop $(P\bar P)$ & $a$ & $b_0$ & $b_+$\\
        \midrule
        \multirow{2}{*}{$K^+$} & $\pi^+\pi^-$ & $-\frac12(1-r_\pi^2)$ & $0$ & $1$\\
         & $K^+K^-$ & $+\frac12(1-r_\pi^2)$ & $0$ & $1$\\
        \midrule
        \multirow{2}{*}{$\frac1{\sqrt{2}}K^0$} & $\pi^+\pi^-$ & $\frac12{r_\pi^2}$ & $1$ & $0$\\
         & $K^+K^-$ & $-\frac12(3-r_\pi^2)$ & $-2$ & $-1$\\
    \bottomrule
    \end{tabular}
    \caption{
        On-shell $K\pi PP$ amplitudes at LO in ChPT in terms of the coefficients of ansatz \eqref{eq:KpiPP_LO_gen}.
        All the coefficients are in the units of $\go$.
        Note the additional factor of $\sqrt{2}$ that should multiply all the coefficients of the neutral-channel entries.
    }
    \label{tab:amplitudes}
\end{table}

\section{Lorentz structure of the $K\to\pi\gamma^*\gamma^*$ amplitude}
\label{sec:structures}

The amplitude for the $K\to\pi\gamma^*\gamma^*$ vertex can be separated into independently gauge-invariant parts based on multiple Lorentz structures, related to how two EM tensors contract with (up to) two derivatives.
To NLO [i.e., $\mathcal{O}(p^6)$] up to which it is common and sensible to calculate these processes, the two-photon gauge-invariant tensor structures can be represented by a general expression
\begin{multline}
M_K^4T_{\rho\sigma}^{(4)}(k_1,k_2;p_1,p_2)
\equiv
k_1\cdot p_1k_2\cdot p_2g_{\rho\sigma}
+k_1\cdot k_2p_{1\rho}p_{2\sigma}
\\
-k_1\cdot p_1k_{2\rho}p_{2\sigma}
-k_2\cdot p_2p_{1\rho}k_{1\sigma}
\,.
\end{multline}
By construction, for any 4-momenta $p_1$ and $p_2$,
\begin{equation}
\begin{aligned}
k_1^\rho T_{\rho\sigma}^{(4)}(k_1,k_2;p_1,p_2)&=0\,,\\
k_2^\sigma T_{\rho\sigma}^{(4)}(k_1,k_2;p_1,p_2)&=0\,.
\end{aligned}
\end{equation}
From now on, I will keep the photon momenta implicit, i.e., $T_{\rho\sigma}^{(4)}(p_1,p_2)\equiv T_{\rho\sigma}^{(4)}(k_1,k_2;p_1,p_2)$.
Notice also that $T_{\rho\sigma}^{(4)}(p_1,p_2)$ is linear in both arguments.

There are several other things to notice.
First, the LO structure $T_{\rho\sigma}^{(2)}$ defined by
\begin{equation}
\begin{aligned}
M_K^2T_{\rho\sigma}^{(2)}(k_1,k_2)
&\equiv k_1\cdot k_2g_{\rho\sigma}-k_{1\sigma}k_{2\rho}\\
&=\frac{M_K^4}{k_1\cdot k_2}\,T_{\rho\sigma}^{(4)}(k_2,k_1)
\end{aligned}
\label{eq:T(2)rs}
\end{equation}
can be reproduced in this notation, too [cf.~\cref{eq:structs} and the shorthand notation $\hat T_{\rho\sigma}^{(1)}\equiv T_{\rho\sigma}^{(2)}(k_1,k_2)$].
Even more generally,
\begin{equation}
\begin{aligned}
T_{\rho\sigma}^{(2)}(k_1,k_2)
=\frac{M_K^2}{k_1\cdot p}\,T_{\rho\sigma}^{(4)}(p,k_1)
=\frac{M_K^2}{k_2\cdot p}\,T_{\rho\sigma}^{(4)}(k_2,p)\,,
\end{aligned}
\end{equation}
for any $p$.
Second, one can easily reproduce the structures of the amplitudes used in the literature for the doubly on-shell and singly off-shell cases.
Concentrating on the nonanomalous parts,
\begin{multline}
    \mathcal{M}^{(2a)}_{\rho\sigma}\big(K(P)\to\pi(r)\gamma_\rho(k_1)\gamma_\sigma(k_2)\big)\\
    =-A\,M_K^2T_{\rho\sigma}^{(2)}(k_1,k_2)+B\,\frac{M_K^2}{k_1\cdot k_2}T_{\rho\sigma}^{(4)}(P,P)
\end{multline}
is the same as Eq.~(2) in Ref.~\cite{Gabbiani:1998tj}, and, similarly,
\begin{multline}
    \mathcal{M}^{(2a)}_{\rho\sigma}\big(K(P)\to\pi(r)\gamma_\rho^*(k_1)\gamma_\sigma(k_2)\big)\\
    =-\widetilde A\,M_K^2T_{\rho\sigma}^{(2)}(k_1,k_2)
    +\widetilde B\,\frac{M_K^2}{k_1\cdot k_2}T_{\rho\sigma}^{(4)}(P,P)\\
    +\widetilde D\,\frac{M_K^2}{k_1\cdot k_2}T_{\rho\sigma}^{(4)}(k_1,P)
\end{multline}
is Eq.~(23) therein (leaving out the explicit factor $\frac\alpha{2\pi}$).
Finally, $\hat T_{\rho\sigma}^{(2)}$ of \cref{eq:structs} can be written as
\begin{equation}
    \hat T_{\rho\sigma}^{(2)}
    =\hat T_{\rho\sigma}^{(1)}+\frac{4M_K^2k_1\cdot k_2}{\lambda_s}\big[T_{\rho\sigma}^{(4)}(k_1,k_2)-T_{\rho\sigma}^{(4)}(k_2,k_1)\big]\,,
\end{equation}
or inverting this for $T_{\rho\sigma}^{(4)}(k_1,k_2)$,
\begin{equation}
    T_{\rho\sigma}^{(4)}(k_1,k_2)
    =\frac1{M_K^2k_1\cdot k_2}\Big(
        k_1^2k_2^2\hat T_{\rho\sigma}^{(1)}+\tfrac14\lambda_s\hat T_{\rho\sigma}^{(2)}
    \Big)\,,
\end{equation}
and $\widetilde F_{\rho\sigma}$ of \cref{eq:Ftilde_rs} is
\begin{multline}
    \widetilde F_{\rho\sigma}\bigl(k_1;k_2\bigr)
    =F(k_1^2)\,\frac{M_K^4}{{(2P\cdot k_2-k_2^2)(2r\cdot k_2+k_2^2)}}\\
    \qquad\times\big[(M_K^2-M_\pi^2)T_{\rho\sigma}^{(4)}(k_1,2r+k_2)\\
    \qquad\qquad-k_1^2T_{\rho\sigma}^{(4)}(2r+k_1,2r+k_2)\big]\,.
\label{eq:Ftilde_rs_T4}
\end{multline}

\section{Counterterms}
\label{sec:CTs}

\subsection{Counterterms at LO}

In this section, I discuss $\mathcal{O}(p^4)$ (LO) CTs.

For $K\pi\gamma(\gamma)$, there is only one independent monomial per each channel, for instance (see also \cref{sec:EFT} for more details),
\begin{equation}
    \partial^\mu K^+F_{\mu\nu}\partial^\nu\pi^-\,,\qquad
    \partial^\mu K^0F_{\mu\nu}\partial^\nu\pi^0\,,
\label{eq:L_CT_g(g)_terms}
\end{equation}
and their hermitian conjugates.
Similarly, for $K\pi\gamma\gamma$, two independent terms can be written down:
\begin{equation}
    K^+F_{\mu\nu}F^{\mu\nu}\pi^-\,,\qquad
    K^0F_{\mu\nu}F^{\mu\nu}\pi^0\,.
\label{eq:L_CT_gg_terms}
\end{equation}
However, the two-photon contact interaction for the neutral channel is only allowed by chiral symmetry at higher orders, and the second term in \cref{eq:L_CT_gg_terms} does not follow from the chiral Lagrangian \eqref{eq:L_ChPT}.
In what follows, I will use a channel-neutral notation collectively for the pairs of the terms above,
\begin{equation}
    \Lb_\text{CT}^{(4)}
    \supset ie\GF c^{(4)}D^\mu KF_{\mu\nu}D^\nu\pi
    +e^2\GF d^{(4)}KF_{\mu\nu}F^{\mu\nu}\pi\,,
\label{eq:L_2_CT}
\end{equation}
with, in general, different coefficients for the charged and neutral channels.
The corresponding amplitudes are
\begin{align}
&\begin{aligned}
    &\frac1{e\GF}\mathcal{M}_{\text{CT},\,\rho}^{(4)}\big(K(P)\to\pi(r)\gamma_\rho^*(k)\big)\\
    &\qquad=-c^{(4)}(k^2r_\rho-k\cdot rk_\rho)\,,
    \label{eq:M4_CT_r}
\end{aligned}\\[2mm]
&\begin{aligned}
    &\frac1{e^2\GF}\mathcal{M}_{\text{CT},\,\rho\sigma}^{(4)}\big(K(P)\to\pi(r)\gamma_\rho^*(k_1)\gamma_\sigma^*(k_2)\big)\\
    &\qquad=-4d^{(4)}M_K^2T_{\rho\sigma}^{(2)}(k_1,k_2)\\
    &\qquad+\xi c^{(4)}
    \big[k_{1\sigma}(k_1+k_2)_\rho-g_{\rho\sigma}k_1\cdot(k_1+k_2)\\
    &\qquad\qquad\qquad\qquad+\{k_1\leftrightarrow k_2,\rho\leftrightarrow\sigma\}\big]\,,
    \label{eq:M4_CT_rs}
\end{aligned}
\end{align}
where the term proportional to $\xi$ only contributes in the charged channel (for which $\xi=+1$), whereas it vanishes for the neutral channel ($\xi=0$), in which the covariant derivative in \cref{eq:L_2_CT} equals the partial one.

In the case of LO ChPT, Lagrangian~\eqref{eq:L_ChPT} leads to
\begin{alignat}{2}
    c_+^{(4)}&=\go\frac23(12L_9-w_1-2w_2)\,,\quad &d_+^{(4)}&=\go\frac23(6L_{10}-w_4)\,,\label{eq:c/d(4)+}\\
    c_0^{(4)}&=\go\frac{\sqrt{2}}3(w_1-w_2+3w_3)\,, &d_0^{(4)}&=0\,.\label{eq:c/d(4)0}
\end{alignat}
The only nonvanishing CT contribution in the two-photon case at LO in ChPT thus arises in the charged channel, as the neutral channel lacks the 1PR part as well as any of the two direct contributions of \cref{eq:M4_CT_rs} and the 1PI part must thus be UV-finite.

In the charged channel, when the 1PR part (in the sense of \cref{fig:K->pigg}) is subtracted from \cref{eq:M4_CT_rs}, the remainder from $\mathcal{M}_{\text{CT},\,\rho\sigma}^{(4)}$ is
\begin{equation}
    -e^2\GF M_K^2\big[4d_+^{(4)}+2c_+^{(4)}\big]T_{\rho\sigma}^{(2)}(k_1,k_2)\,.
\end{equation}
In ChPT at LO, this is
\begin{equation}
    -\frac43e^2G_8M_K^2\big[12(L_9+L_{10})-w_1-2w_2-2w_4\big]T_{\rho\sigma}^{(2)}(k_1,k_2)\,,
\end{equation}
which acts as the (finite) CT for the 1PI part of the charged-channel amplitude [cf.~\cref{eq:c_hat}, and how it enters \cref{eq:Kpigg_LO}].

\subsection{Counterterms for $K\pi\gamma(\gamma)$ at NLO}

The CTs discussed in this section are instrumental for achieving a finite $K\to\pi\gamma^*$ amplitude at $\mathcal{O}(p^6)$ (NLO).
I again treat both channels in parallel.

Regarding the three-particle $K(P)\pi(r)\gamma(k)$ vertex, in view of the 4-momentum conservation $P+r+k=0$, there are four independent kinematical invariants at $\mathcal{O}\bigl(p^2\bigr)$: $M_K^2$, $k^2$, $r^2$, and $k\cdot r$.
Consequently, there are four independent gauge-invariant structures at $\mathcal{O}\bigl(p^6\bigr)$ (see also second paragraph of \cref{sec:EFT}) stemming, for instance, from the following Lagrangian monomials:
\begin{align}
    M_K^2\partial^\mu KF_{\mu\nu}\partial^\nu\pi\,,\quad
    \partial^\mu K\partial_\alpha\partial^\alpha F_{\mu\nu}\partial^\nu\pi\,,\notag\\
    \partial_\alpha\partial^\alpha K\partial^\mu F_{\mu\nu}\partial^\nu\pi\,,\quad
    \partial^\mu K\partial^\nu F_{\mu\nu}\partial_\alpha\partial^\alpha\pi\,.
\end{align}
Working at a specific order, the last two monomials become dynamically redundant utilizing equations of motion;
they also lead to terms proportional to $(P^2-M_K^2)$ and $(r^2-M_\pi^2)$ that disappear on-shell (plus terms like those from the next section in the charged case).
Note that even after plugging these into 1PR diagrams (in the sense of \cref{fig:K->pigg}), the final result for the $K^+\to\pi^+\gamma^*\gamma^*$ amplitude does not depend on such off-shell extensions.
The relevant Lagrangian then features two terms, related to two $\mathcal{O}\bigl(p^2\bigr)$ independent kinematical invariants on-shell,
\begin{multline}
\Lb_\text{CT}^{(6)}
\supset\frac{ie\GF}{\ChPT^2}\Big[
\cA_1 M_K^2\partial^\mu KF_{\mu\nu}\partial^\nu\pi\\
+\cA_2\partial^\mu K\partial_\alpha\partial^\alpha F_{\mu\nu}\partial^\nu\pi
\Big]\,.
\label{eq:L_4_CT_g(g)}
\end{multline}
These represent my choice of CT basis for $K\pi\gamma$.
Similar to the previous subsection, employing $\partial_\mu\to D_\mu$ (which only has an effect for the charged channel) gives rise to associated $K^+\pi^-\gamma\gamma$ vertices, which combined with the 1PR part (in the sense of \cref{fig:K->pigg}) guarantee the gauge invariance of the $K^+\to\pi^+\gamma^*\gamma^*$ amplitude.
(There are, of course, no 1PR contributions in the sense of \cref{fig:K->pigg} in the neutral channel.)

In analog with \cref{eq:M4_CT_r,eq:M4_CT_rs}, the above CT Lagrangian \eqref{eq:L_4_CT_g(g)} leads to the following matrix elements:
\begin{align}
&\begin{aligned}
&\mathcal{M}_{\text{CT},\,\rho}^{(6)}\big(K(P)\to\pi(r)\gamma_\rho^*(k)\big)\\
&=-e\GF(k^2r_\rho-k\cdot rk_\rho)\frac1{\ChPT^2}\big[\cA_1M_K^2-\cA_2k^2\big]\,,
\label{eq:M6_CT_r}
\end{aligned}\\[2mm]
&\begin{aligned}
    &\mathcal{M}_{\text{CT},\,\rho\sigma}^{(6)}
    \big(K(P)\to\pi(r)\gamma_\rho^*(k_1)\gamma_\sigma^*(k_2)\big)\\
    &=\xi e^2\GF\frac1{\ChPT^2}\big[\cA_1M_K^2-\cA_2k_1^2\big]\\
    &\qquad\times\big[k_{1\sigma}(k_1+k_2)_\rho-g_{\rho\sigma}k_1\cdot(k_1+k_2)\big]\\
    &\qquad\qquad\qquad\qquad+\{k_1\leftrightarrow k_2,\rho\leftrightarrow\sigma\}\,.
    \label{eq:M6_CT_rs}
\end{aligned}
\end{align}

\subsection{Counterterms for $K\pi\gamma\gamma$ at NLO}

In this section, the set of CT required to make the $K\to\pi\gamma^*\gamma^*$ amplitude at NLO finite is presented.
One shall look for independent gauge-invariant structures at $\mathcal{O}\bigl(p^6\bigr)$, i.e., with two electromagnetic-field tensors and two derivatives.
I arrived at six terms and chose the basis leading to the Lagrangian
\begin{align}
\begin{split}
&\Lb_\text{CT}^{(6)}
\supset\frac{e^2\GF}{\ChPT^2}\frac14\Big[
\cB_1 M_K^2KF_{\mu\nu}F^{\mu\nu}\pi\\
&+\cB_2 \partial_\alpha KF_{\mu\nu}F^{\mu\nu}\partial^\alpha\pi
+\cB_3 K\partial_\alpha F_{\mu\nu}\partial^\alpha F^{\mu\nu}\pi\\
&+2\cB_4 F^{\mu\alpha}F_{\alpha\nu}(\pi\partial_\mu\partial^\nu K+K\partial_\mu\partial^\nu\pi)\\
&+2\cB_5 F^{\mu\alpha}F_{\alpha\nu}(\pi\partial_\mu\partial^\nu K-K\partial_\mu\partial^\nu\pi)\\
&+2\cB_6 K\partial_\mu F^{\mu\nu}\partial^\alpha F_{\alpha\nu}\pi
\Big]\,.
\end{split}
\label{eq:L_6_CT_gg}
\end{align}
Following~\eqref{eq:L_6_CT_gg}, the amplitude for the $K\to\pi\gamma\gamma$ vertex can be written as a linear combination of independent gauge-invariant Lorentz structures.
I will use the notation introduced in \cref{sec:structures} and express the result in terms of a single object $T_{\rho\sigma}^{(4)}(p_1,p_2)$, while employing \mbox{$T_{\rho\sigma}^{(2)}(k_1,k_2)=\frac{M_K^2}{k_1\cdot k_2}T_{\rho\sigma}^{(4)}(k_2,k_1)$} as in \cref{eq:T(2)rs}.

Regarding the terms proportional to $T_{\rho\sigma}^{(2)}(k_1,k_2)$, one has to accompany it with all independent [$\mathcal{O}(p^2)$] scalar products.
For a four-body process (like a three-body decay), one has seven independent structures: three scalar products minus one kinematical constraint beyond linear 4-momentum conservation, four 4-momenta squared, and $M_K^2$.
Bose symmetry $\{k_1\leftrightarrow k_2,\,\rho\leftrightarrow\sigma\}$ adds additional restrictions.
Adding the $\mathcal{O}(p^4)$ gauge-invariant structures, there are, in addition, three linearly independent Bose-symmetric combinations.
In particular, the CT amplitude for the $K\to\pi\gamma^*\gamma^*$ vertex based (solely) on \eqref{eq:L_6_CT_gg} is
\begin{align}
\begin{aligned}
&\frac1{e^2\GF M_K^2}\frac{\ChPT^2}{M_K^2}\mathcal{M}_{\text{CT},\,\rho\sigma}^{(6)}\bigl(K(P)\to\pi(r)\gamma^*(k_1)\gamma^*(k_2)\bigr)\\
&=\frac1{M_K^2}\bigl[-\cB_1M_K^2-\cB_2P\cdot r+\cB_3k_1\cdot k_2
\big]T_{\rho\sigma}^{(2)}(k_1,k_2)\\
&-\cB_4\big[T_{\rho\sigma}^{(4)}(P,P)+T_{\rho\sigma}^{(4)}(r,r)\big]\\
&-\cB_5\big[T_{\rho\sigma}^{(4)}(P,P)-T_{\rho\sigma}^{(4)}(r,r)\big]
+\cB_6T_{\rho\sigma}^{(4)}(k_1,k_2)\,.
\end{aligned}
\label{eq:M_CT(6)}
\end{align}
Thus, there are two extra terms with respect to the related literature:
In the doubly off-shell case, one should consider the terms proportional to $\cB_5$ and $\cB_6$.

Alternatively, the term proportional to $\cB_4$ could be replaced by $\partial_\mu KF^{\mu\alpha}F_{\alpha\nu}\partial^\nu\pi$, which leads to the structure
\begin{equation}
\begin{aligned}
&T_{\rho\sigma}^{(4)}(P,r)+T_{\rho\sigma}^{(4)}(r,P)\\
&=\big[T_{\rho\sigma}^{(4)}(P,P)+T_{\rho\sigma}^{(4)}(r,r)\big]
-T_{\rho\sigma}^{(4)}(k_1,k_2)\\
&+\frac1{M_K^2}(2P\cdot r+k_1\cdot k_2-P^2-r^2)T_{\rho\sigma}^{(2)}(k_1,k_2)\,.
\end{aligned}
\label{eq:T4_Pr+rP}
\end{equation}
This shows the relation of the two $\cB_4$ terms, which simplifies when at least one of the photons is on-shell (and the $\cB_6$ term does not contribute).
From \cref{eq:T4_Pr+rP}, it also follows that the tensor structure of the UV-divergent part shown in \cref{eq:Mrs_1PI_UV} can be simply written as
\begin{equation}
    \big[T_{\rho\sigma}^{(4)}(P,r)+T_{\rho\sigma}^{(4)}(r,P)\big]-\frac{P\cdot r}{M_K^2}\,T_{\rho\sigma}^{(2)}(k_1,k_2)\,.
\end{equation}
Finally, the term proportional to $\cB_5$ can be written as
\begin{multline}
    T_{\rho\sigma}^{(4)}(P,P)-T_{\rho\sigma}^{(4)}(r,r)\\
    =\frac12\big[T_{\rho\sigma}^{(4)}(k_1,P+r)+T_{\rho\sigma}^{(4)}(P+r,k_2)\big]\\
    +\frac12\frac1{M_K^2}(P^2-r^2)T_{\rho\sigma}^{(2)}(k_1,k_2)\,,
\end{multline}
which is utilized in \cref{eq:Kpigg_pol}.

\section{Effective Lagrangian for the 1PR part}
\label{sec:EFT}

The goal of this appendix is to explore ways of writing down a gauge-invariant effective Lagrangian describing the $K^+\to\pi^+\gamma(\gamma)$ transitions in a situation when the one-photon transition is extracted from data.
The building blocks are thus (up to two) electromagnetic field-strength tensor(s) $F^{\mu\nu}$ and the kaon [$K=K(x)$] and pion [$\pi=\pi(x)$] fields, which need to be decorated with appropriate powers of covariant derivatives $D_\mu=\partial_\mu-i\eta eA_\mu$, with $\eta$ being the charge sign of the field on which it acts (here, $\eta=\pm1\text{ or }0)$.

At lowest order, one starts with the term $K^+F^{\mu\nu}\pi^-$, adding $D_\mu$ and $D_\nu$ to obtain a Lorentz scalar, identifying independent terms leading to the $K^+(P)\to\pi^+(r)\gamma(k)$ amplitude.
This is directly connected to scalar products that can be built from 4-momenta $P$, $r$, and $k$, and form gauge-invariant Lorentz structures.
Having $2+1$ (two $D$'s and one $F$) momenta, this can in general be written as $(k\cdot p\,q_{\rho}-k\cdot q\,p_\rho)$, with $p$ and $q$ being any of the three particle momenta.
For a nonvanishing result, $p$ and $q$ cannot be the same, and with the 4-momentum conservation $P=k+r$ at the vertex, there is only one independent option, and I choose, for instance, $(k^2r_\rho-k\cdot r\,k_\rho)$.
The related interaction term is $K^+\partial_\mu F^{\mu\nu}\partial_\nu \pi^-$.
Promoting the partial derivatives to covariant, one needs to add an extra term for a nonvanishing derivative commutator, which is the same thing as when one would consider a term with two photons directly, i.e., $K^+F_{\mu\nu}F^{\mu\nu}\pi^-$.
Finally, the case of two photons allows for introducing another independent Lorentz-invariant structure: $\varepsilon_{\mu\nu\rho\sigma}K^+F^{\mu\nu}F^{\rho\sigma}\pi^-$.
This is driven by one-particle-exchange contributions, dominantly given by $\pi^0$ or $\eta$ poles, and it will be left out from the following discussion and treated separately elsewhere.

Working directly with the mentioned term $K^+F^{\mu\nu}\pi^-$ and two additional derivatives on the Lagrangian level, the same is achieved in the following manner.
Out of nine possible combinations (leaving out, for simplicity, the meson charges and spacetime coordinates),
\begin{equation}
\begin{matrix}
(D_\mu D_\nu K)F^{\mu\nu}\pi,&
D_\nu KD_\mu F^{\mu\nu}\pi,&
D_\nu KF^{\mu\nu}D_\mu\pi,\\
D_\mu KD_\nu F^{\mu\nu}\pi,&
K(D_\mu D_\nu F^{\mu\nu})\pi,&
KD_\nu F^{\mu\nu}D_\mu\pi,\\
D_\mu KF^{\mu\nu}D_\nu\pi,&
KD_\mu F^{\mu\nu}D_\nu\pi,&
KF^{\mu\nu}(D_\mu D_\nu\pi)
\,,
\end{matrix}
\label{eq:DDKFpi}
\end{equation}
only two turn out dynamically independent.
To show this, one might first use the identity $D_\mu D_\nu=\frac12\{D_\mu,D_\nu\}+\frac12[D_\mu,D_\nu]$, followed by $[D_\mu,D_\nu]=-i\eta eF_{\mu\nu}$.
The ``diagonal'' terms then give rise to $-\frac12i\eta eK^+F_{\mu\nu}F^{\mu\nu}\pi^-$.
Next, note that the terms ``above the diagonal'' are equivalent to the terms ``under'' simply by renaming $\mu\leftrightarrow\nu$ and using the antisymmetry of $F^{\mu\nu}$.
There are thus only four unique nonvanishing terms in total present in \cref{eq:DDKFpi}, and I can, for instance, eliminate the whole third line.
By construction, the terms in each line add up to a total derivative due to the vanishing net charge of each term, i.e., e.g.,
\begin{equation}
D_\mu(D_\nu K^+F^{\mu\nu}\pi^-)
=\partial_\mu(D_\nu K^+F^{\mu\nu}\pi^-)\,,
\end{equation}
and thus dynamically to zero.
Thus, one of the three respective terms in each line (for instance the last term) can be expressed as a linear combination of the other two.
In particular, from the first line, $D_\nu K^+F^{\mu\nu}D_\mu\pi^-$ can be expressed as a linear combination of $D_\nu K^+D_\mu F^{\mu\nu}\pi^-$ and $K^+F_{\mu\nu}F^{\mu\nu}\pi^-$, and the second line says an equivalent statement about $K^+D_\nu F^{\mu\nu}D_\mu\pi^-$, as $D_\mu K^+D_\nu F^{\mu\nu}\pi^-=-D_\nu K^+D_\mu F^{\mu\nu}\pi^-$.
I thus end up with two independent terms, for instance, $D_\mu K^+\partial_\nu F^{\mu\nu}\pi^-$ and $K^+F_{\mu\nu}F^{\mu\nu}\pi^-$ (chosen on purpose different from but equivalent to the first approach).

To extend the Lagrangian so that it can generate matrix elements including a realistic form factor, one can add more partial derivatives to the term $K^+\partial_\mu F^{\mu\nu}\partial_\nu\pi^-$.
They can only be added in pairs, leading to scalar products $P^2$, $r^2$, $k^2$, $P\cdot r$, $k\cdot P$, and $k\cdot r$.
All these can be written (on-shell and with $P=k+r$) as linear combinations of masses squared and $k^2$; hence, the form factor $F(k^2)$ is the function of $k^2$ only.
This formally allows for introducing an operator
\begin{equation}
\w\equiv\sum_{n=0}^\infty\frac{\hat c_n}{n!}\bigg[\frac{-\partial_\alpha\partial^\alpha}{\ChPT^2}\bigg]^{n}
\end{equation}
added to the Lagrangian in the following way:
\begin{equation}
\begin{aligned}
\Lb
&=ie\GF\partial_\mu K^+(\w\,\partial_\nu F^{\mu\nu})\pi^-\\
&\simeq ie\GF K^+(\w\,\partial_\mu F^{\mu\nu})\partial_\nu\pi^-\\
&\simeq-ie\GF\partial_\mu K^+(\w F^{\mu\nu})\partial_\nu\pi^-\,.
\end{aligned}
\end{equation}
The complex coefficients $\hat c_n$ are chosen in such a way that
\begin{equation}
F(k^2)=\GF\sum_{n=0}^\infty\frac{\hat c_n}{n!}\biggl(\frac{k^2}{\ChPT^2}\biggr)^{n}\,,
\end{equation}
or, explicitly,
\begin{equation}
    \hat c_n
    =\frac1\GF\frac{\text{d}^n}{\text{d}z^n}{F(zM_K^2)}\Big|_{z=0}\,,\qquad
    n\in \mathbb{N}\,.
\end{equation}
Promoting partial derivatives to covariant, one arrives at
\begin{multline}
\Lb
=ie\GF D_\mu K^+(\w\,\partial_\nu F^{\mu\nu})\pi^-\\
-\hat\kappa e^2\GF\frac14K^+F_{\mu\nu}F^{\mu\nu}\pi^-\,.
\label{eq:LKpigg}
\end{multline}

Starting with the first term,
\begin{equation}
    \Lb
    \supset ie\GF D_\mu K^+(\w\,\partial_\nu F^{\mu\nu})\pi^-\,,
\end{equation}
the discussion proceeds along the same lines as at other places in this paper.
This term generates both the standard matrix element for the $K^+\to\pi^+\gamma^*$ transition,
\begin{multline}
\mathcal{M}_\rho\big(K^+(P)\to\pi^+(r)\gamma_\rho^*(k)\big)\\
=eF(k^2)
\big[
k^2P_\rho
-(k\cdot P)k_\rho
\big]
\label{eq:K->pig}
\end{multline}
[cf.~\cref{eq:M_K->Pg}], and it also leads to the term complementing the 1PR part of the $K^+\to\pi^+\gamma^*\gamma^*$ amplitude [the one driven by the one-photon transition, proportional to $F(k^2)$]:
\begin{equation}
\begin{split}
&\mathcal{M}_{\rho\sigma}^{(b)}\big(K^+(P)\to\pi^+(r)\gamma_\rho^*(k_1)\gamma_\sigma^*(k_2)\big)\\
&=e^2F(k_1^2)\big(k_1^2g_{\rho\sigma}-k_{1\rho}k_{1\sigma}\big)+\{k_1\leftrightarrow k_2\}\,.
\end{split}
\label{eq:M_Kpigg_b}
\end{equation}
Plugging thus the general $K^+\to\pi^+\gamma^*$ vertex from \cref{eq:K->pig} into the 1PR diagrams (the first two) of \cref{fig:K->pigg} and combining this with $\mathcal{M}_{\rho\sigma}^{(b)}$ of \cref{eq:M_Kpigg_b} leads to a gauge-invariant structure
\begin{equation}
\begin{aligned}
&\mathcal{M}_{\rho\sigma}^{\text{1PR}}\big(K^+(P)\to\pi^+(r)\gamma_\rho^*(k_1)\gamma_\sigma^*(k_2)\big)\\
&=e^2F(k_1^2)\bigg\{
(k_1^2r_\rho-r\cdot k_1 k_{1\rho})\,\frac{(2P-k_2)_\sigma}{2P\cdot k_2-k_2^2}\\
&\qquad-(k_1^2P_\rho-P\cdot k_1 k_{1\rho})\,\frac{(2r+k_2)_\sigma}{2r\cdot k_2+k_2^2}\\
&\qquad+\big(k_1^2g_{\rho\sigma}-k_{1\rho}k_{1\sigma}\big)
\bigg\}
+\{k_1\leftrightarrow k_2,\rho\leftrightarrow\sigma\}\,,
\end{aligned}
\end{equation}
or, in the language of \cref{eq:F_rs,eq:Ftilde_rs},
\begin{equation}
\begin{aligned}
\mathcal{M}_{\rho\sigma}^{\text{1PR}}
=e^2F_{\rho\sigma}\bigl(k_1,k_2\bigr)\,.
\end{aligned}
\end{equation}
Note that in case one of the photons is on-shell ($k_2^2=0$, $k_{2\sigma}[\epsilon^\sigma(k_2)]\to0$) and the other is contracted with the conserved current ($k_{1\rho}\to0$), as is the case of $K^+\to\pi^+\ell^+\ell^-\gamma$ decays, one simply finds
\begin{equation}
\begin{split}
&\mathcal{M}_{\rho\sigma}^\text{1PR}\big(K^+(P)\to\pi^+(r)\gamma_\rho^*(k_1)\gamma_\sigma(k_2)\big)
\Big|_{\substack{\scalebox{0.8}{$\displaystyle k_{1\rho}\to\,0$}}}\\
&=e^2\widetilde F_{\rho\sigma}\bigl(k_1;k_2\bigr)
\Big|_{\substack{\scalebox{0.8}{$\displaystyle k_2^2\,\to\,0,\,k_{2\sigma}[\epsilon^\sigma(k_2)]\to\,0,\,k_{1\rho}\to\,0$}}}\\
&=e^2k_1^2F(k_1^2)
\biggl(r_\rho\frac{P_\sigma}{P\cdot k_2}
-P_\rho\frac{r_\sigma}{r\cdot k_2}
+g_{\rho\sigma}\biggr)\,,
\end{split}
\end{equation}
with $\widetilde F_{\rho\sigma}\bigl(k_1;k_2\bigr)$ from \cref{eq:Ftilde_rs}.
For both photons on-shell, $\mathcal{M}_{\rho\sigma}^\text{1PR}=0$.

The previous treatment would not be complete without having the second monomial of \cref{eq:LKpigg},
\begin{equation}
    \Lb
    \supset-\hat\kappa e^2\GF\frac14K^+F_{\mu\nu}F^{\mu\nu}\pi^-\,,
\label{eq:L2_kappa}
\end{equation}
which leads to
\begin{multline}
\mathcal{M}_{\rho\sigma}^{(\hat\kappa)}\big(K^+(P)\to\pi^+(r)\gamma_\rho^*(k_1)\gamma_\sigma^*(k_2)\big)\\
=\hat\kappa e^2\GF\big[(k_1\cdot k_2)g_{\rho\sigma}-k_{1\sigma}k_{2\rho}\big]\,.
\label{eq:K->pigg_kappa}
\end{multline}
This part of the amplitude is manifestly gauge invariant on its own.
At lowest order in the discussed EFT, the prefactor ($\hat\kappa$) is just a constant.
This can be compared to its manifestation in the LO ChPT calculation \eqref{eq:Kpigg_LO}, and it would relate to the CT contribution, $\hat\kappa=-2\go\kappa\hat c$, or to $\big(\tfrac1{\GF M_K^2}\big)G(k_1^2,k_2^2)$ of \cref{eq:Kpigg_NLO}.
The 1PI part of the LO ChPT result~\eqref{eq:Kpigg_LO} then expands on this simple amplitude $\mathcal{M}_{\rho\sigma}^{(\hat\kappa)}$ in terms of $A_\xi^{(1)}$ and by including an additional form factor $A_\xi^{(2)}$.

\section{Explicit expressions}
\label{sec:Ai}

In this appendix, I list the explicit expressions for the subamplitudes entering \cref{eq:Kpigg_NLO}.
For compactness, I am using slightly different notation than in the rest of the paper.
First, the subamplitudes $A_0^{(i)}$ and $A_1^{(i)}$ defined below relate to those used in \cref{eq:Kpigg_NLO,eq:Kpigg_NLO_A1A2} in the following way:
\begin{gather}
    A_{c_+}^{(i)}=A_0^{(i)}+A_1^{(i)}(z_1,z_2,y)+A_1^{(i)}(z_2,z_1,-y)\,,\; i=1,2\,,\notag\\
\begin{aligned}
    A_{c_+}^{(3)}=A_0^{(3)}&+A_1^{(3)}(z_1,z_2)+A_1^{(3)}(z_2,z_1)\,,\\
    A_{c_+}^{(4)}(k_1,k_2)&=A^{(4)}(z_1,z_2,y)\,,\\
    A_{c_+}^{(4)}(k_2,k_1)&=A^{(4)}(z_2,z_1,-y)\,.
\end{aligned}
\end{gather}
The transition subamplitudes depend on photon virtualities,
\begin{equation}
    z_1=\frac{k_1^2}{M_K^2}\,,\qquad
    z_2=\frac{k_2^2}{M_K^2}\,,
\end{equation}
and two independent variables,
\begin{align}
    z&=\frac1{M_K^2}\,(k_1+k_2)^2\,,\notag\\
    y&=\frac1{M_K^2}\,(k_1-k_2)\cdot(P+r)\,.
\end{align}
The subamplitudes $A_0^{(i)}\equiv A_0^{(i)}(z_1,z_2)=A_0^{(i)}(z_2,z_1)$, $i=1,2,3$, are symmetric under ($z_1\leftrightarrow z_2$), and for simplicity, I omit the arguments.
I further use
\begin{align}
    z_0&=\frac1{M_K^2}\,k_1\cdot k_2\,,\notag\\
    \lambda_z&=\lambda(z,z_1,z_2)=4(z_0^2-z_1z_2)\,,
\end{align}
so $z=z_1+z_2+2z_0$.
Moreover, I introduced
\begin{align}
    r_P&=\frac{M_P}{M_K}\,,\notag\\
    y_0&=\frac1{M_K^2}\,(P+r)^2=2+2r_\pi^2-z\,,\\
    y_{1,2}&=\frac1{M_K^2}\,k_{1,2}\cdot(P+r)
    =\frac12(1-r_\pi^2\pm y)\,.\notag
\end{align}
It follows that $y_1+y_2=1-r_\pi^2$ and $y_1-y_2=y$.
For compactness of the expressions, I also employed
\begin{align}
    z_*&=z_1z_2\,,\notag\\
    z_{ij}&=z_i+z_j\,,\notag\\
    z_3&=zz_1z_2\,,\\
    z_P&=z_3+\lambda_P\,,\notag\\
    \lambda_P&=r_P^2\lambda_z\,.\notag
\end{align}
Regarding the loop functions, $B_0(x)=6r_P^2\frac1x\Jbar_P(M_K^2x)$ and $C_0=C_0(M_K^2z,M_K^2z_1,M_K^2z_2;M_P^2)$.

With the above choices, the subamplitudes read
\begin{widetext}
\begin{align*}
&A_0^{(1)}
=\frac{1}{18z_0\lambda_P\lambda_z}
\Bigl\{
18M_P^2C_0z_P\bigl\{\lambda_P(zz_0-\tfrac14\lambda_z)-\tfrac{1}{2}y_0z_0z_P
+z\bigl[z_3z_0-\tfrac14\lambda_z(z_*+2zz_0)\bigr]\bigr\}\\
&+B_0(z)zz_0\bigl\{5zz_{01}z_{02}\lambda_P
-\tfrac12y_0\bigl[\tfrac12z\lambda_z^2+z^2z_0(3z_*-\tfrac12\lambda_z)+\lambda_P(5zz_0-2\lambda_z)\bigr]
+z^2\bigl[z_*(3zz_0-\tfrac34\lambda_z)+\lambda_z(\tfrac78\lambda_z-2zz_0)\bigr]\bigr\}\\
& +\lambda_P\Bigl(9z_{01}z_{02}\lambda_P-\tfrac12y_0z_0\bigl\{9z_P-2z_0[z_*+(3z-z_0)z_0]\bigr\}+3z_*z^2z_0+\lambda_z\bigl[\tfrac94z_3-6z^2z_0+\tfrac18\lambda_z(9z_{12}+11z_0)\bigr]\Bigr)\\
& +\frac{4y_1y_2}{\lambda_z}
\Bigl[
18M_P^2C_0z_P(z_P+4zz_0^2-z_0\lambda_z)(z_*+2z_0^2)
+3\lambda_P\bigl\{15z_*^2z+\lambda_z\bigl[z_*(4z-\tfrac32z_0)+\tfrac14\lambda_z(z+z_0)\bigr]+3\lambda_P(z_*+2z_0^2)\bigr\}\\
& \qquad +B_0(z)z_0\Bigl(z\bigl\{
9z_3[5z_3+\lambda_z(z-z_0)]
+\tfrac14\lambda_z^2\bigl(z^2-\lambda_z\bigr)\bigr\}
+\lambda_P\bigl[39z^2z_*+\lambda_zz(\tfrac{13}2z_{12}+7z_0)-\tfrac12\lambda_z^2\bigr]\Bigr)
\Bigr]
\Bigr\}
\,,
\stepcounter{equation}\tag{\theequation}
\end{align*}
\begin{align*}
&A_1^{(1)}(z_1,z_2,y)
=\frac{1}{36z_0\lambda_P\lambda_z}
\Bigl[
y_0B_0(z_1)z_1^2z_{02}z_0\bigl[z_1(3z_2z-\tfrac12\lambda_z)+5\lambda_P\bigr]\\
& \quad -\frac{8y_1y_2}{\lambda_z}B_0(z_1)z_1\bigl\{z_1[16z_0^6+48z_0^5z_2+4z_0^4(z_{02}z_1+6z_2^2)+4z_0^3z_*(12z_{02}+7z_1)+2z_0^2z_*z_2(9z_{02}+35z_1)\\
& \quad +z_0z_*^2(13z_{12}+8z_2)+3z_*^2z_2(z_2-z_1)]
+\lambda_P[39z_*(z_*+z_1z_0)+\tfrac12\lambda_z(25z_*+13z_1z_0)+\tfrac12\lambda_z^2]\bigr\}\\
& -2B_0(z_1)z_1z_{02}\bigl\{3z_*^2z_1(z_2-z_1)+z_*z_0z_1\bigl(11z_{12}^2-2z_2^2\bigr)+z_0^2z_*(33z_*+49z_1^2)\\
& \quad -z_0^3\bigl(8z_1^3-24z_*z_1+6z_*z_2\bigr)
-z_0^4(34z_1^2+24z_*)-28z_1z_0^5+\lambda_P\bigl[4z_0^3-3z_1^2z_2+z_0(5z_1^2+13z_1z_0+z_*)\bigr]\bigr\}\\
& +\frac1{\lambda_z}(1-r_\pi^2+y)^2\Bigl(-18M_P^2C_0z_0z_2z_P\bigl[-z_1^2z_2+4z_0^2(3z_2+4z_0)+z_1(z_2+2z_0)(3z_2+8z_0)+3\lambda_P\bigr]\\
& \quad +3z_0\lambda_P\bigl[-15z_*z_2z-\tfrac12\lambda_z(5z_*+3z_2^2+5z_2z_0)+\tfrac12\lambda_z^2-9z_2\lambda_P\bigr]\\
& \quad -B_0(z)z_0\bigl\{z\bigl[16z_*^2z_2^2+z_*z_0z_2\bigl(19z_1^2+39z_2^2+90z_*\bigr)+4z_*z_0^2z_2(39z_{12}-2z_2)+2z_0^3z_2\bigl(17z_1^2+3z_2^2+90z_*\bigr)\\
& \quad\quad +8z_0^4z_2(3z_{12}-z_2)-8z_1z_0^5\bigr]\\
& \quad\quad +\lambda_P\bigl[8z_*z_2(3z_{12}-z_2)+z_0z_2\bigl(31z_{12}^2+40z_*+8z_2^2\bigr)+4z_0^2z_2(27z_{12}+8z_2)+4z_0^3(2z_{12}+31z_2)+24z_0^4\bigr]\bigr\}\\
& \quad +B_0(z_1)z_1z_0\bigl\{z_1\bigl[3z_*z_2(z_*+3z_2^2)+z_0z_*(19z_*+51z_2^2)+12z_0^2z_2(9z_*+3z_2^2)+2z_0^3(17z_*+42z_2^2)+24z_2z_0^4-8z_0^5\bigr]\\
& \quad +\lambda_P\bigl[39z_*z_{02}+2\lambda_z(z_{02}+2z_2)\bigr]\bigr\}\\
& \quad +B_0(z_2)z_2^2z_0\bigl\{z_2\bigl[7z_*^2-3z_1^2z_*+39zz_*z_0+z_0^2(48z_1^2+34z_*)+z_0^3(96z_1+6z_2)+16z_0^4\bigr]
+\lambda_P\bigl(39z_2z_{01}+8\lambda_z\bigr)\bigr\}\Bigr)
\Bigr]\,,
\stepcounter{equation}\tag{\theequation}
\end{align*}
\begin{align*}
&A_0^{(2)}
=\frac{1}{36z_0\lambda_P\lambda_z}
\Bigl(
9M_P^2C_0\bigl[\lambda_z(z_P+2zz_0z_{01})(z_P+2zz_0z_{02})
+2y_0z_0z_P(z_P+4zz_0^2-z_0\lambda_z)\bigr]\\
& +B_0(z)z\Bigl(y_0z_0\bigl\{3z_3(5zz_0-\lambda_z)+\tfrac14z\lambda_z(2zz_0-\lambda_z)+\lambda_P(13zz_0-2\lambda_z)\bigr\}
+\tfrac12zz_0\lambda_z[5\lambda_P+17z_3+2zz_0(9z-7z_0)]\Bigr)\\
&+3y_0z_0\lambda_P\bigl(3z_P-\tfrac12z_0\lambda_z+2zz_0^2\bigr)+\tfrac12\lambda_P\lambda_z\bigl[9\lambda_P+3z_3+18z^2z_0+\tfrac12\lambda_z(z_0-3z_{12})\bigr]\\
& +\frac{y_1y_2}{\lambda_z}\Bigl\{
-{12\lambda_P\bigl[6\lambda_P(z_*+2z_0^2)+70zz_*^2+\lambda_zz_*(28z_{12}+43z_0)+\lambda_z^2(3z_{12}+4z_0)\bigr]}\\
& \quad -8B_0(z)z_0\bigl\{z
\bigl[105z_3^2+5\lambda_zz_3(7z_{12}+8z_0)+\tfrac14\lambda_z^2\bigl(12z^2-13zz_0+6z_*\bigr)+\tfrac18\lambda_z^3\bigr]
+\lambda_P\bigl[55zz_3+\lambda_zz(\tfrac{21}{2}z-8z_0)-\tfrac12\lambda_z^2\bigr]\bigr\}\\
& \quad -144M_P^2C_0\Bigl[\lambda_P^2(3z_*+\tfrac12\lambda_z)+2\lambda_P\{z_*(z+2z_0)(5z_*+\lambda_z)+10z_0^4(z-z_0)\}
+\tfrac18\lambda_z^3\bigl(z^2+z_*-zz_0\bigr)\\
& \quad \quad
+\lambda_z^2z_*\bigl(\tfrac{11}{4}z^2-z_*-\tfrac{11}{4}zz_0\bigr)
+\lambda_zz_*^2\bigl(\tfrac{3}{2}z_{12}^2-12z_*-4zz_0+16z^2\bigr)
+z_*^3\bigl(29z^2-24z_*+24zz_0+6z_{12}^2\bigr)
\Bigr]\Bigr\}
\Bigr)
\,,
\stepcounter{equation}\tag{\theequation}
\end{align*}
\begin{align*}
&A_1^{(2)}(z_1,z_2,y)
=\frac{1}{36z_0\lambda_P\lambda_z}
\Bigl\{
\frac{8y_1y_2}{\lambda_z}B_0(z_1)z_1\bigl\{
4z_0^4\bigl(15z_*z_2+6z_2z_0^2+12z_0^3\bigr)+z_*^2z_1^2(13z_0-3z_2)\\
& \qquad +z_*z_1^2\bigl(3z_2^3+21z_2^2z_0+102z_2z_0^2+44z_0^3\bigr)+6z_1^2z_0^2\bigl(3z_2^3+21z_2^2z_0+8z_0^3\bigr)+2z_1(50z_*z_0^4+60z_2z_0^5+58z_0^6)\\
& \qquad +\lambda_P\lambda_z^2+z_1\lambda_P\bigl[55z_*z_{02}+\tfrac12\lambda_z(37z_2+21z_0)\bigr]
\bigr\}\\
& +B_0(z_1)\Bigl(-y_0z_1z_0\bigl\{z_1\bigl[z_1z_2^2(5z_1+3z_2)+(13z_1+21z_2)z_*z_0+12z_2(3z_1+z_2)z_0^2+2(z_1+12z_2)z_0^3+4z_0^4\bigr]\\
& \qquad +\lambda_P[5z_*+z_0(13z_1+8z_0)]\bigr\}
-\tfrac12\lambda_z\bigl\{z_1\bigl[3z_*^2(z_2-z_1)+z_0z_*\bigl(11z_1^2+6z_2^2+25z_*\bigr)+z_0^2\bigl(18z_1^3+80z_1z_*+42z_1z_2^2\bigr)\\
& \qquad +2z_0^3\bigl(41z_1^2+6z_2^2+56z_*\bigr)+16z_0^4(7z_1+3z_2)+52z_0^5\bigr]
+\lambda_P[-4z_0^3+4z_1z_0(z_2+2z_0)+z_1^2(-3z_2+5z_0)]\bigr\}\Bigr)\\
& +\frac{z_0}{4\lambda_z}(1-r_\pi^2+y)^2
\Bigl[12\lambda_P\bigl\{35z_*\bigl[z_*+z_2(z_2+2z_0)\bigr]+\tfrac12z_2\lambda_z(23z_{01}+13z_2)+\tfrac12\lambda_z^2+9z_2\lambda_P\bigr\}\\
& \qquad +4B_0(z)\bigl\{z\bigl[
16z_*^2z_2^2+z_*z_0z_2\bigl(3z_1^2+55z_2^2+90z_*\bigr)+8z_*z_0^2z_2(21z_1+31z_2)+2z_0^3z_2\bigl(47z_1^2+25z_2^2+210z_*\bigr)\\
& \qquad \qquad +4z_0^4z_2(61z_1+39z_2)+8z_0^5(z_1+15z_2)+8z_0^6
\bigr]\\
& \qquad \qquad +\lambda_P\bigl[
8z_*(3z_*+2z_2^2)+z_0z_2\bigl(39z_1^2+55z_2^2+126z_*\bigr)+12z_0^2(13z_*+17z_2^2)+4z_0^3(4z_1+51z_2)+40z_0^4\bigr]\bigr\}\\
& \qquad -4B_0(z_1)\bigl\{z_1\bigl[3z_1^2z_2^3(z_1+3z_2)+3(z_1+21z_2)z_*^2z_0+12z_1z_2^2(11z_1+6z_2)z_0^2+2(47z_1+102z_2)z_*z_0^3\\
& \qquad \qquad +4z_2(43z_1+6z_2)z_0^4+8(z_1+6z_2)z_0^5+8z_0^6\bigr]
+\lambda_P[-8z_0^4+16z_1z_0^2(3z_2+z_0)+3z_1^2z_2(5z_2+13z_0)]\bigr\}\\
& \qquad -4B_0(z_2)z_2^2\bigl[(-3z_1+7z_2)z_*^2+z_1z_2^2(27z_1+55z_2)z_0+4(9z_1+44z_2)z_*z_0^2+2z_2(108z_1+25z_2)z_0^3\\
& \qquad \qquad +12(6z_1+11z_2)z_0^4+72z_0^5+7z_*\lambda_P+z_0\lambda_P(55z_2+48z_0)\bigr]\\
& \qquad +72M_P^2C_0\Bigl(8z_0^4\lambda_P+3z_2\lambda_P^2
+z_2[-z_1^4z_2^2+8z_{02}z_0^4(z_2+2z_0)+2z_1^3z_2(z_2^2+4z_2z_0+6z_0^2)\\
& \qquad \qquad +3z_1^2(z_2+2z_0)(z_2^3+6z_2^2z_0+12z_2z_0^2+4z_0^3)+4z_1z_0^2(6z_2^3+23z_2^2z_0+30z_2z_0^2+12z_0^3)]\\
& \qquad \qquad +2z_2\lambda_P[z_*(11z_{12}+4z_2+30z_0)+\tfrac12\lambda_z(5z_{12}+z_2+9z_0)]\Bigr)\Bigr]
\Bigr\}\,,
\stepcounter{equation}\tag{\theequation}
\end{align*}
\begin{gather}
A_0^{(3)}
=\frac1{9\lambda_P\lambda_z}
\Bigl[
\tfrac{1}{2}\lambda_P\lambda_z(3z_{12}+z_0)-3\lambda_P\bigl(3\lambda_P+z_3\bigr)
+B_0(z)z^2z_0\bigl(2zz_0^2-5z_P\bigr)
-18M_P^2C_0z_P^2
\Bigr]\,,\\
A_1^{(3)}(z_1,z_2)
=\frac1{9\lambda_P\lambda_z}z_1B_0(z_1)
\Bigl\{
\lambda_P\bigl(5z_{01}z_0+\tfrac34\lambda_z\bigr)
+3z_*^2(z_2-z_1)
+z_1\bigl[z_*z_0(9z_{12}-4z_1)+20z_*z_0^2-2z_1z_0^3-8z_0^4\bigr]
\Bigr\}\,,
\end{gather}
\begin{align*}
&A_4(z_1,z_2,y)
=\frac1{9\lambda_P\lambda_z^2}\Bigl[(1-r_\pi^2+y)\Bigl(
-3\lambda_P\bigl[3z_2\lambda_P+3z_*^2+2z_0^2z_{02}(z_2+2z_0)+z_*(3z_2^2+4z_2z_0-2z_0^2)\bigr]\\
& -B_0(z)zz_0\bigl\{z\bigl[15z_*z_2z+\tfrac12\lambda_z(6z_*+z_2z)+\tfrac14\lambda_z^2\bigr]+\lambda_P\bigl(8z_{02}^2+5z_2z\bigr)\bigr\}
+B_0(z_1)\bigl\{\lambda_P\bigl[13z_*z_1z_{02}+2\lambda_zz_1z_{02}-\tfrac12\lambda_z^2\bigr]\\
& \quad +z_1\bigl[3z_1^2z_2^4-3z_1z_2^3(z_1-4z_0)z_{01}+4z_0^5(z_1+2z_0)+2z_*z_0^3(3z_1+2z_0)+z_1z_2^2z_0(z_{01}+5z_0)(5z_{01}+z_0)\bigr]
\bigr\}\\
& +B_0(z_2)z_2^2\bigl\{3z_*^2(z_1-z_2+3z_0)+13z_*z_2^2z_0+40z_*z_2z_0^2+24z_*z_0^3+2z_0^3\bigl(z_2^2+4z_2z_0+6z_0z_{01}\bigr)+\lambda_P\bigl[\tfrac34\lambda_z+13z_0z_{02}\bigr]\bigr\}\\
& \quad -18M_P^2C_0\bigl\{zz_*z_2\bigl(z_3+2\lambda_P\bigr)+4z_0^2z_{02}^2z_P+z_2\lambda_P^2\bigr\}\Bigr)\\
& +(1-r_\pi^2-y)z_0\Bigl(
3\lambda_P\bigl[3\lambda_P+5z_3+\tfrac12\lambda_z(z-z_0)\bigr]
+B_0(z)z\bigl\{z\bigl[15z_3z_0+\lambda_z(\tfrac12zz_0-3z_*)-\tfrac14\lambda_z^2\bigr]+\lambda_P\bigl(13zz_0-2\lambda_z\bigr)\bigr\}\\
& \quad -B_0(z_1)z_1A_4^{(z_1)}(z_1,z_2)
-B_0(z_2)z_2A_4^{(z_1)}(z_2,z_1)
+18M_P^2C_0z_P\bigl[\lambda_P+5z_3+\lambda_z(z-z_0)\bigr]
\Bigr)
\Bigr]\,,
\stepcounter{equation}\tag{\theequation}
\end{align*}
with
\begin{equation}
    A_4^{(z_1)}(z_1,z_2)
    =z_1\bigl[15z_*\bigl(z_2^2+z_0z_1+3z_2(z_0+z_1)\bigr)+\tfrac12\lambda_z\bigl(22z_*+6z_2^2+z_0z_1+12z_0z_2\bigr)+\tfrac14\lambda_z^2\bigr]
    +\lambda_P\bigl(5z_*+13z_1z_0+8z_0^2\bigr)\,.
\end{equation}
\end{widetext}


\let\raggedright

\providecommand{\href}[2]{#2}\begingroup\raggedright\endgroup

\end{document}